\documentclass[journal,twoside,web]{IEEEtran}

\usepackage{cite}

\usepackage{amsmath,amssymb,amsfonts,amsthm}

\allowdisplaybreaks
\usepackage{enumitem}

\usepackage{graphicx}

\usepackage{hyperref}
\usepackage{subfigure}
\usepackage{romannum}

\usepackage{array}

\usepackage{ulem}
\usepackage{color}

\usepackage{nomencl}
\makenomenclature

\usepackage{algorithm}
\usepackage[noend]{algpseudocode}

\renewcommand{\nomlabel}[1]{\hbox to 1.35cm{#1\hfill}}
\newcommand{\hangpara}[1]{%
	\parbox[t]{7.3cm}{\hangindent=0em #1}
}

\theoremstyle{remark}
\newtheorem{definition}{Definition}
\newtheorem{example}{Example}
\newtheorem{remark}{Remark}
\newtheorem{problem}{Problem}
\newtheorem{theorem}{Theorem}
\newtheorem{lemma}{Lemma}

\hypersetup{hidelinks=true}
\usepackage{textcomp}
\def\BibTeX{{\rm B\kern-.05em{\sc i\kern-.025em b}\kern-.08em
    T\kern-.1667em\lower.7ex\hbox{E}\kern-.125emX}}
\begin{document}
	\pagenumbering{arabic}
\title{Global and Local Error-Tolerant Decentralized State Estimation Under Partially Ordered Observations}
\author{Dajiang Sun, Christoforos N. Hadjicostis, \IEEEmembership{Fellow, IEEE}, and Zhiwu Li, \IEEEmembership{Fellow, IEEE}
	\thanks{Dajiang Sun is with the School of Electro-Mechanical Engineering, Xidian University, Xi'an, 710071, China (e-mail: djsun@stu.xidian.edu.cn).}%
	\thanks{Christoforos N. Hadjicostis is with the Department of ECE, University of Cyprus, Nicosia, Cyprus (e-mail: hadjicostis.christoforos@ucy.ac.cy).}
	\thanks{Zhiwu Li is with the Institute of Systems Engineering, Macau University of Science and Technology, Taipa, Macau SAR, China, and also with the School of Electro-Mechanical Engineering, Xidian University, Xi'an, 710071, China (e-mail: zhwli@xidian.edu.cn).}}

\maketitle

\begin{abstract}
We investigate decentralized state estimation for a discrete event system in a setting where the information received at a coordinator may be corrupted or tampered by a malicious attacker. Specifically, a system is observed by a set of (local) observation sites (OSs) which occasionally send their recorded sequences of observations to the coordinator that is in charge of estimating the system state. The malfunctions and attacks, referred to as errors in this paper, include symbol deletions, insertions and replacements, each of which bears a positive cost. Two types of errors, global errors and local errors, are proposed to describe the impact of errors on decentralized information processing. Global errors occur when all OSs record the same error, while local errors occur when different OSs record different errors. Distinguishing these types of errors is important for a proper design of decentralized information processing (so as to be more resilient and better equipped to handle attacks and failures). For each type of error, we propose two methods to efficiently perform state estimation: one based on appropriately modifying the original system and the other based on inferring the matching behavior of the original system. For each method, we adopt an \textit{estimation-by-release} methodology to design an algorithm for constructing a corresponding synchronizer for state estimation.
\end{abstract}

\nomenclature[01]{$\mathbb{Z}_{\geq 0}$}{Set of non-negative integers.}
\nomenclature[02]{$\mathcal{I}$}{$\{1,\dots,m\}$, index set of OSs.}
\nomenclature[03]{$O_i$}{$i$-th observation site.}
\nomenclature[04]{$\Sigma_i$}{Set of events observed by $O_i$.}
\nomenclature[05]{$\Sigma_{\mathcal{I}}$}{$\bigcup_{i\in\mathcal{I}}\Sigma_i$, set of events observed by at least one OS.}
\nomenclature[06]{$c_u$}{Cost constraint.}
\nomenclature[07]{$\mathbf{C_u}$}{$\{0,1,...,c_u\}$, set of costs corresponding to $c_u$.}
\nomenclature[08]{$[R]^{c_u}$}{Error Relation Matrix (ERM) with $c_u$.}
\nomenclature[09]{$\overline{\mathcal{R}}^{c_u}(\cdot)$}{Set of erroneous sequences of events with costs.}
\nomenclature[10]{$\tau_g$}{Erroneous SI-state under global error (E$_g$SI-state).}
\nomenclature[11]{$G_g$}{Cost-constrained globally modified system.}
\nomenclature[12]{$\mathcal{S}_g(\cdot,\cdot)$}{E$_g$-synchronizer.}
\nomenclature[13]{$\mathcal{TO}^{c_u}_g(\cdot)$}{\hangpara{Set of global-error-tolerant TO-sequences (GETO-sequences) upon ``$\cdot$''.}}
\nomenclature[14]{$\widetilde{B}_g$}{\hangpara{Global-error-tolerant-sequence-builder (E$_g$TS-builder).}}
\nomenclature[15]{$\mathcal{\widetilde{S}}_g(\cdot,\cdot)$}{E$_g$T-synchronizer.}
\nomenclature[16]{$\{[R_i]\}^{c_u}_{i\in\mathcal{I}}$}{Set of local ERMs with cost constraint $c_u$.}
\nomenclature[17]{$\tau_l$}{Erroneous SI-state under local error (E$_l$SI-state).}
\nomenclature[18]{$G_l$}{Cost-constrained locally modified system.}
\nomenclature[19]{$B^m$}{Multi-S-builder (MS-builder).}
\nomenclature[20]{$\mathcal{S}_l(\cdot,\cdot)$}{E$_l$-synchronizer.}
\nomenclature[21]{$\mathcal{TO}^{c_u}_l(\cdot)$}{\hangpara{Set of local-error-tolerant TO-sequences (LETO-sequences) upon ``$\cdot$''.}}
\nomenclature[22]{$\widetilde{B}_l$}{\hangpara{Local-error-tolerant-sequence-builder (E$_l$TS-builder).}}
\nomenclature[23]{$\mathcal{\widetilde{S}}_l(\cdot,\cdot)$}{E$_l$T-synchronizer.}

\printnomenclature

\section{Introduction}
\IEEEPARstart{W}{ith} the proliferation of information and communication technologies, large-scale systems such as cyber-physical systems have become increasingly complex and interconnected. These systems comprise several interacting components and are typically geographically distributed, which can be advantageous for fault tolerance, safety, and security. During the operation of these systems, a primary objective is to utilize information from sensors to estimate the system's state.

In the context of discrete event systems (DESs), characterized by dynamic systems with discrete-state spaces and event-driven mechanisms, the state estimation problem \cite{Hadjicostis2020,LinWangHanShen2020,LaiLahayeGiua2019,SearRudie2014} aims to identify a set of possible states that necessarily includes the true state of the system, considering various models and observation settings. 
This task serves as the foundation for various applications, including control \cite{RamadgeWonham2008, LinWonham1988}, diagnosis \cite{Lin1994,Sampath1995,DeboukLafortuneTeneketzis2003}, security and privacy \cite{SabooriHadjicostis2007,Sweeney2002}, and synthesis problems \cite{YinLafortune2016a, YinLafortune2016b}. 
The presence of variations in information processing settings and possibly inaccurate observations makes this task challenging. 
In \cite{LinWangHanShen2020}, the authors present two methods to achieve conservative and exact state estimates in a multichannel networked DES, also considering communication delays. 
The work in \cite{LaiLahayeGiua2019} focuses on I-detectability and I-opacity, which are directly related to initial-state estimation in DES modeled by unambiguous weighted automata.

Recently, approaches to state estimation have evolved to include possibly erroneous observations \cite{LiHadjicostisWuLi2023,Yin2017,CarvalhoWuKwongLafortune2018,Su2018,MeiraMarchandLafortune2019,WakaikiTabuadaHespanha2019,AmmourAmariBrennerDemongodinLefebvre2023,Zhang2022}. 
These involve abnormalities like cyber attacks \cite{LimaAlvesCarvalhoMoreira2022,OliveiraLealTeixeiraLopes2023,YuGaoCongWuSong2023}, malfunctioning sensors and actuators \cite{GoesKangKwongLafortune2020,TaiLinSu2023,HeWuLi2024}, and communication errors \cite{FritzZhang2023,GheitasiLucia}. 
In \cite{LiHadjicostisWuLi2023}, the authors tackle centralized state estimation in the presence of a bounded number of adversarial attacks. 
Additionally, the work in \cite{Yin2017} introduces corresponding notions in a stochastic setting by considering probabilistic sensor failures.  
Carvalho \textit{et al.} \cite{CarvalhoWuKwongLafortune2018} address the problem of intrusion detection and mitigation in supervisory control systems, where the actuators can be attacked. 
In \cite{Su2018}, a supervisor synthesis approach is proposed to defy the cyber attack problem, using an automaton to describe an attack model that can intercept and alter sensor measurements. 
The work in \cite{MeiraMarchandLafortune2019} formulates the problem of synthesizing a supervisor capable of ensuring robustness against sensor deception attacks, and reduces this problem to a supervisory control problem with arbitrary control patterns. 
Readers are referred to \cite{DuoZhouAbusorrah2022} for a more extensive survey about attacks on cyber-physical systems modeled by time-driven and event-driven models.

In this paper, we consider a DES modeled by a nondeterministic finite automaton, observed by $m$ OSs \cite{DeboukLafortuneTeneketzis2000, Tripakis2004}, where each OS records and reports the sequence of events it observes.
Specifically, during the system's evolution, an OS determines, based on its own observations, whether the sequence of events it has observed should be sent to the coordinator.
If an OS decides to transmit its observation sequence, it signals its intention to the coordinator.
The coordinator then decides whether to request from all OSs their sequences of observations, referred to as partially ordered sequences of observations (PO-sequences).
The process of requesting and analyzing the information from OSs, to determine the possible system states, is termed \textit{synchronization}.
The specific time instants at which \textit{synchronization} is initiated are governed by a predefined rule set within both the OSs and the coordinator, collectively referred to as a \textit{synchronization strategy.}
We refer to the type of information transmitted by OSs, i.e., PO-sequences, as well as the rules adopted by the coordinator to process PO-sequences during each synchronization, as the decentralized observation-based information processing (DO-based protocol).

Under the DO-based protocol, system information is initially captured by sensors and then conveyed through the communication channels to the OSs where it is recorded.
Subsequently, the information is forwarded, via another set of communication channels, to the coordinator at synchronization.
Within each stage, there are potential vulnerabilities where malfunctions or malicious attacks could occur.
For simplicity, we will refer to ``errors'' instead of attacks and malfunctions in the remainder of this work.
In this paper, we assume that tampering can only happen before information is recorded at the OSs, ensuring that the sequences received by the coordinator are identical to those recorded by the OSs. 
However, the specific stage at which information tampering occurs (prior to being recorded at the OSs) and the type of events that are tampered with remain critical considerations.
If errors occur during the transmission of sequences from sensors to the OSs, the same information may be recorded differently by various OSs, leading to inconsistencies.
In another scenario, for the events observed by at least two OSs, if the OSs rely on shared sensors, an error will affect the recorded information of all OSs that have access to that sensor. 
Compared with the situation where each OS has its own sensors for shared-observable events, this scenario is still worth studying. 
The decentralized observation setting is a compromise due to the naturally distributed physical components of the system. 
The existence of OSs can serve as a bridge between the system and the coordinator, making it both expensive and unnecessary to assign to each OS its own sensor devices.

This paper concentrates on the decentralized state estimation problem for a DES under a DO-based protocol, where local and global errors may occur. 
A global error describes a scenario where the occurrence of an error is recorded (identically) by all relevant OSs; a local error is different in the sense that the occurrence of an error affects the information recorded at only one of the OSs. 
As a result, different PO-sequences received at the coordinator might be subject to different errors.
The errors are formalized as event deletions, insertions, and replacements.  
A deletion or replacement error involves deleting or altering certain data in the information flow. 
An insertion error occurs when non-existent data is deliberately or unintentionally inserted into the information flow. 
Intuitively, the information flow in the system is processed in the following manner. 
First, a sequence of events is generated by the system. 
Then, this sequence is projected into PO-sequences, recorded at different OSs. 
Finally, these PO-sequences are sent to the coordinator for further processing. 
With this understanding in mind, ``global errors'' refer to those errors that occur in the information flow {\em before} being projected into PO-sequences and recorded by the OSs; while ``local errors'' refer to those errors that may occur in each PO-sequence {\em before} they get recorded by different OSs.

The problem of current-state estimation under the DO-based protocol (DO-CSE) was initially proposed in \cite{Hadjicostis2016Partial}.
The primary challenge in performing DO-CSE is to systematically infer the possible order of events occurring in different PO-sequences, effectively reconstructing all potential sequences of the system's observations. 
In \cite{Hadjicostis2016Partial}, a partial-order-based method was introduced to accomplish this task recursively, under the condition that the sets of observable events for all OSs are disjoint.
In our recent work \cite{SunHadjicostisLi2023}, we relax this constraint by introducing the concept of the sequence-builder (\textit{S-builder}), which serves as the foundation for an algorithm designed to estimate system states using a breadth-first search approach. 
When considering the errors, not only do we need to determine the possible order of events showing in PO-sequences, but also simultaneously account for the potential errors within these sequences. 
This dual consideration introduces a higher level of complexity, making the estimation process substantially more challenging than in error-free scenarios.
Especially when local errors occur, inconsistencies may arise among the sequences provided by different OSs. 
For example, an observable event that is deleted in one PO-sequence might be replaced in another.
In this paper, we approach the problems from two perspectives:

1) Modifying the original system: 
This involves re-modeling the system to incorporate potential errors, ensuring the system behaves consistently with recorded PO-sequences.

2) Inferring possible original system sequences: 
This approach aims to infer the sequences of original observations (prior to tampering) without altering the system model.

We generalize our approach in \cite{SunHadjicostisLi2023} and develop a methodology called \textit{estimation-by-release}.
This methodology enables us to progressively estimate the possible states of the system in order, as events are processed (called \textit{release} procedure). 
By determining the order of events and updating the system's state simultaneously, we reduce computational complexity and improve the efficiency of the state estimation process.
The main contributions of this paper are as follows.
 First, it formulates an error model with a matrix based on the error relation.
Specifically, the entries of the matrix represent the possible errors associated with costs. 
Second, the problems of global/local error-tolerant decentralized state estimation are defined under the corresponding error model. 
Third, for each error type, two methods are presented to accomplish the state estimation task: one method is based on modifying the system and the other method is based on inferring the matching sequences. 
We adopt an \textit{estimation-by-release} methodology to design the corresponding algorithms in order to reduce computational complexity, which is analyzed explicitly.

\section{Preliminaries}
\subsection{System Model}
We adopt conventional notations for a DES from \cite{Hadjicostis2020,CassandrasLafortune2008}. Let $\Sigma$ be a finite set of events.  $\Sigma^*$ denotes the set of all finite-length sequences over $\Sigma$, including the empty string $\epsilon$. A language $L\subseteq\Sigma^*$ is a set of strings. Given strings $s,t\in\Sigma^*$, the concatenation of strings $s$ and $t$ is the string $st$ (or $s\cdot t$), i.e., the sequence of symbols in $s$ followed by the sequence of symbols in $t$.  Also, the concatenation of two languages $L_1$ and $L_2$ is defined by $L_1L_2=\{s_1s_2|s_1\in L_1,s_2\in L_2\}$. We say $\sigma\in s$ if there exist $w,t\in \Sigma^*$, $\sigma\in\Sigma$, such that $s=w\sigma t$. The length of a string $s$, denoted by $|s|$, is the number of symbols (events) in $s$, with $|\epsilon|=0$. We use $s/t$ to denote the symbol sequence after $t$ in $s$ (note that $s/s=\epsilon$ and $s/\epsilon=s$). For any string $s\in\Sigma^*$, we define $s\cdot\epsilon=\epsilon\cdot s=s$.

The system considered in this paper is modeled by a nondeterministic finite automaton $G=(Q,\Sigma, \delta, Q_0)$, where $Q$ is the finite set of states, $\Sigma$ is the finite set of events, $\delta: Q\times\Sigma\rightarrow 2^Q$ is the next-state transition function, and $Q_0\subseteq Q$ is the set of initial states. The event $\sigma\in\Sigma$ is said to be defined at $q\in Q$ if $|\delta(q,\sigma)|>0$. The domain of transition function $\delta$ can be extended from $Q\times\Sigma$ to $Q\times\Sigma^*$ in the standard recursive manner: $\delta(q, \sigma s)=\delta(\delta(q, \sigma), s)=\bigcup_{q'\in\delta(q,\sigma)}\delta(q',s)$ for any $q\in Q$, $\sigma\in\Sigma$, and $s\in\Sigma^*$. The behavior of $G$ from state $q\in Q$ is the language $L(G,q)=\{s\in\Sigma^*| \delta(q,s)\neq\emptyset\}$. If there exists a set of marked states $Q_m$ with $Q_m\subseteq Q$, the marked behavior of $G$ is $L_m(G)=\{s\in\Sigma^*|\exists q_0\in Q_0,\delta(q_0,s)\cap Q_m\neq\emptyset\}$.

\subsection{Decentralized Observation-Based Information Processing}
Under decentralized observation-based information processing, we assume that a system is observed by $m$ observation sites (OSs) whose index set is denoted by $\mathcal{I}=\{1,\dots,m\}$. For any OS $O_i$, $i\in\mathcal{I}$, we use $\Sigma_{o,i}$ to denote the set of events that can be observed by $O_i$. Then, the natural projection function $P_{\Sigma_{o,i}}: \Sigma^*\rightarrow\Sigma^*_{o,i}$ is defined recursively as 
\begin{align*}
	P_{\Sigma_{o,i}}(\epsilon)=\epsilon 
	\quad 
	\text{and}
	\quad
	P_{\Sigma_{o,i}}(s\sigma)=
	\begin{cases}
		P_{\Sigma_{o,i}}(s)\sigma, &\text{if} \;  \sigma\in \Sigma_{o,i},\\
		P_{\Sigma_{o,i}}(s),& \text{if} \; \sigma\notin \Sigma_{o,i}.
	\end{cases}
\end{align*}

The natural projection function can map any system behavior to the sequence of observations associated with it. 
For the sake of notational simplicity, $\Sigma_{o,i}$ and $P_{\Sigma_{o,i}}$ will be denoted by $\Sigma_i$ and $P_i$, respectively. 
For any language $L\subseteq \Sigma^*$, define $P_i(L)=\{\omega\in\Sigma^*_i|\exists s\in L, P_i(s)=\omega\}$.
We use notation $P_{\mathcal{I}}$ to denote the natural projection function with respect to (w.r.t.) $\Sigma_{\mathcal{I}}=\bigcup_{i\in\mathcal{I}}\Sigma_{i}$, which is the set of events observed by at least one OS.
The DO-based protocol in this paper is as follows:

1) Each $O_i$ uses its projection function $P_i:L(G,Q)\rightarrow\Sigma_i^*$ to record partially ordered sequences of observations (PO-sequences).
Recorded sequences are reset after each synchronization.

2) Synchronization strategy: 
A specific synchronization strategy\footnote{For example, an OS requests synchronization whenever it observes two events since the last time the coordinator initiated synchronization, while the coordinator initiates synchronization whenever at least one OS requests it.} is not defined in this work, as we focus on state estimation within a single synchronization step.\footnote{This is because, regardless of the strategy performed by the system, the coordinator always receives a set of sequences of observations.}

3) Synchronization function:
At each synchronization, the coordinator updates its current-state estimates based on the received PO-sequences using the synchronization function 
\begin{align*}
\mathcal{S}:\Sigma_1^*\times\dots\times\Sigma_m^*\times 2^Q\rightarrow 2^Q,
\end{align*}
such that for $(\tau^{(1)}, \dots, \tau^{(i)},\dots, \tau^{(m)})\in \Sigma_1^*\times\dots\times \Sigma_i^*\times \dots\times\Sigma_m^*$, $Q'\in 2^Q$, we have
\begin{multline*}
\mathcal{S}(\tau^{(1)}, \dots, \tau^{(i)},\dots, \tau^{(m)},Q')=\{q\in Q|\exists q'\in Q', \\
\exists u\in L(G,q'),\forall i\in \mathcal{I}:P_i(u)=\tau^{(i)}\land q\in\delta(q',u)\}.
\end{multline*}
Here $\tau^{(i)}$ is the sequence of observations sent by $O_i$, with $\tau^{(i)}\in \Sigma^*_{i}$, $i\in \mathcal{I}$, and 
$Q'$ is the set of state estimates obtained at the coordinator at the latest synchronization.
This function also defines DO-based current-state estimation (DO-CSE) \cite{SunHadjicostisLi2023}.

Intuitively, the goal at each synchronization is for the coordinator to fuse and analyze PO-sequences to update in real time its knowledge of the system state.
Suppose that a sequence of events $t$ is executed by a system. 
The synchronization information state (SI-state) is defined as the information recorded by the $m$ OSs: $IS(t)=(P_1(t),\dots,P_m(t))$. 
We use $T$ to denote the set of SI-states and $\tau=(\tau^{(1)},\dots, \tau^{(m)})\in\Sigma^*_1\times\dots\times\Sigma^*_m$ to denote an element of $T$. 
The totally ordered sequences (TO-sequences) upon $\tau$, denoted by $\mathcal{TO}(\tau)$, are defined as a subset of sequences of events in $\Sigma^*_{\mathcal{I}}$,
\begin{align}\label{aaa}
	\mathcal{TO}(\tau)=\{\omega\in\Sigma^*_{\mathcal{I}}|\forall i\in\mathcal{I}: P_i(\omega)=\tau^{(i)}\}.
\end{align}
We use $T_e=(\epsilon,\dots, \epsilon)$ to denote the SI-state after each synchronization, as the transmission of PO-sequences frees the memory of OSs after each synchronization.

We next describe some operators that will be used in the remainder of this paper.
For event $\sigma\in\Sigma$, we use $I(\sigma)=\{i\in\mathcal{I}|\sigma\in\Sigma_i\}$ to denote the index set of OSs that can observe $\sigma$.  For $i\in I(\sigma)$, $\Omega_i(\sigma)=\{\sigma\omega\in\Sigma_i^*\}$ is the set of sequences of observations that start with event $\sigma$ and may be recorded by $O_i$. From the viewpoint of the coordinator, the unobservable reach of the subset of states $Q'\subseteq Q$ is given by
\begin{align*}
	\operatorname{UR}(Q')=\{q\in Q|\exists q'\in Q', \exists u\in(\Sigma\backslash\Sigma_{\mathcal{I}})^*: q\in\delta(q',u)\},
\end{align*}
whereas the observable reach of the subset of states $Q'\subseteq Q$ under $e\in\Sigma_{\mathcal{I}}$ is given by
$\operatorname{R}_e(Q')=\{q\in Q|\exists q'\in Q': q\in\delta(q',e)\}.$
We define $\operatorname{R}_{\epsilon}(Q'):=\operatorname{UR}(Q').$ 

\textbf{Convention:} 
The related definitions and formulations in this paper are described based on a single synchronization step.
We assume, without loss of generality, that the system is known to be in the set of initial states $Q_0$. 
One can easily extend the proposed approaches to a scenario where several consecutively synchronizations are initiated during the system evolution, each time following the steps described for a single synchronization.
We also assume that the coordinator does not have knowledge of which specific OS (or OSs) initiates (or initiate) the synchronization.\footnote{This information, if known, could be used to improve state estimation but it is not crucial in terms of our methodology\cite{SunHadjicostisLi2023}.}

The following is the notion of sequence-builder (S-builder) which was originally proposed in \cite{SunHadjicostisLi2023}. 
The definition has been slightly modified to account for the fact that we cannot be certain which OSs initiate the synchronization.
\begin{definition}\label{def-S-builder}
	\textit{(S-builder)} Given an SI-state $\tau_0$, an S-builder is a five-tuple transition system $B=(T,\Sigma_s,T_0,T_e,h)$ where
	\begin{itemize}
		\item $T\subseteq \Sigma_1^*\times\dots\times\Sigma_m^*$ is the set of SI-states;
		\item $\Sigma_s=\{\sigma|\exists i\in\mathcal{I}: \sigma\in\tau_0^{(i)}\}\subseteq \Sigma_{\mathcal{I}}$ is the set of events appearing in the initial state $T_0$;
		\item $T_0\in T$ is the initial state with $T_0=\tau_0$;
		\item $T_e\in T$ is the ending state;
		\item$h:T\times \Sigma_s\rightarrow T$ is the event release transition function, which is defined as follows: for any $\tau=(\tau^{(1)},\ldots,\tau^{(m)})\in T$, $\tau'=(\tau'^{(1)},\ldots,\tau'^{(m)})\in T$, and  $\sigma\in\Sigma_s$, it holds:
		\begin{multline*}
			h(\tau,\sigma)=\tau'\Rightarrow\forall i\in I(\sigma),\forall j\in\mathcal{I}/I(\sigma):\\ \tau^{(i)}\in\Omega_i(\sigma)\wedge\tau'^{(i)}=\tau^{(i)}/\sigma\wedge\tau'^{(j)}=\tau^{(j)}.
		\end{multline*}
	\end{itemize}
\end{definition}

The S-builder captures TO-sequences for a given SI-state $\tau_0$ without explicitly enumerating them.
The transition $h$ recursively constructs the state space of the S-builder by releasing the events in $\tau_0$.
Readers can refer to \cite{SunHadjicostisLi2023} for more details.

\section{Error Model and its Impacts on Sequences}

Given an observation sequence $\omega$, it is possible for one or more symbols in $\omega$ to be deleted or replaced with other symbols, or for additional symbols to be inserted into $\omega$ at any position. 
Let $\Sigma_1$ and $\Sigma_2$ be two sets of events. Relation $R\subseteq \Sigma_1\times \Sigma_2$ is a mapping $R:\Sigma_1\rightarrow 2^{\Sigma_2}$ defined by $\sigma_2\in R(\sigma_1)$ if and only if $(\sigma_1,\sigma_2)\in R$ where $\sigma_1\in\Sigma_1$ and $\sigma_2\in\Sigma_2$. 
With this definition at hand, the error relation could be defined as follows.

\begin{definition}
	Given a set of events $\Sigma_{\mathcal{I}}$, the error relation on $\Sigma_{\mathcal{I}}$ is defined as $R_{e,\mathcal{I}}\subseteq ((\Sigma_{\mathcal{I}}\cup\{\epsilon\})\times(\Sigma_{\mathcal{I}}\cup\{\epsilon\}))$, where $(\sigma_1,\sigma_2)\in R_{e,\mathcal{I}}$ indicates that $\sigma_1$ could be altered into $\sigma_2$. More specifically, $R_{e,\mathcal{I}}=R_{d,\mathcal{I}}\dot{\cup}R_{in,\mathcal{I}}\dot{\cup}R_{r,\mathcal{I}}\dot{\cup}R_{no,\mathcal{I}}$ where
	\begin{enumerate}
		\item $R_{d,\mathcal{I}}$ is said to be a deletion (error) relation if $R_{d,\mathcal{I}}=\{(\sigma_1,\sigma_2)|\sigma_1\in\Sigma_\mathcal{I},\sigma_2=\epsilon\}$;
		\item $R_{in,\mathcal{I}}$ is said to be an insertion (error) relation if $R_{in,\mathcal{I}}=\{(\sigma_1,\sigma_2)|\sigma_1=\epsilon, \sigma_2\in\Sigma_\mathcal{I}\}$;
		\item $R_{r,\mathcal{I}}$ is said to be a replacement (error) relation if $R_{r,\mathcal{I}}=\{(\sigma_1,\sigma_2)|\sigma_1, \sigma_2\in\Sigma_{\mathcal{I}}\wedge \sigma_1\neq\sigma_2\}$;
		\item $R_{no,\mathcal{I}}$ is said to be an error-less relation if $R_{no,\mathcal{I}}=\{(\sigma_1,\sigma_2)|\sigma_1= \sigma_2\in\Sigma_{\mathcal{I}}\cup\{\epsilon\}\}$.
	\end{enumerate}
\end{definition}

\begin{remark}
The error relation provides a uniform framework for the (non)occurrence of errors between two specific symbols, $\sigma_1,\sigma_2\in\Sigma_{\mathcal{I}}\cup\{\epsilon\}$. 
Pair $(\sigma_1,\sigma_2)\in R_{in,\mathcal{I}}$ describes the scenario where an event  $\sigma_2$ is inserted in the observation sequences with $\sigma_1=\epsilon$; pair $(\sigma_1,\sigma_2)\in R_{d,\mathcal{I}}$ (respectively, $(\sigma_1,\sigma_2)\in R_{r,\mathcal{I}}$) describes the situation that $\sigma_1$ is deleted (respectively, replaced) in the observation sequence with $\sigma_2=\epsilon$ (respectively, $\sigma_2\in\Sigma_{\mathcal{I}}$, $\sigma_2\neq\sigma_1$). 
Note that, $R_{no,\mathcal{I}}$, a reflexive relation which indicates that no error happens in the specific symbol, is included in $R_{e,\mathcal{I}}$ to allow for a comprehensive definition of the relation between symbols. We say that an ``error action $(\sigma_1,\sigma_2)$'' occurs if $\sigma_1$ is altered into $\sigma_2$ provided that $(\sigma_1,\sigma_2)\in R_{e,\mathcal{I}}$ holds with $\sigma_1\neq\sigma_2$, and that an ``error-less action $(\sigma_1,\sigma_1)$'' occurs when symbol $\sigma_1$ is unaltered in the sequence. 
In the rest of the paper, $R$ will be used instead of $R_{e,\mathcal{I}}$ whenever it is clear from context. 
The inclusion of $(\epsilon,\epsilon) \in R_{no,\mathcal{I}}$ ensures the function's completeness.
\end{remark}

As mentioned in the introductory section, we assume that errors may occur in the observation sequences before they are recorded by the OSs.
We also assume that the errors incurred during one synchronization step have bounded total costs. The definition of cost is as follows: given an error relation $R$, the cost upon $R$ is defined as a function $\mathcal{A}:R\rightarrow \mathbb{Z}_{\geq 0}$ which assigns a cost to an error(-less) action in $R$ (here, $\mathbb{Z}_{\geq 0}$ denotes the set of non-negative integers). Specifically, for all error-less actions $(\sigma_1,\sigma_1)\in R$, it holds that $\mathcal{A}((\sigma_1,\sigma_1))=0$. 
The error relation with cost can be represented by a matrix, denoted by $[R]$, called an \textit{Error Relation Matrix} (ERM), whose row and column indices index the elements of $\Sigma_{\mathcal{I}}\cup\{\epsilon\}$, such that the entries of $[R]$, for all $\sigma_1,\sigma_2\in\Sigma_{\mathcal{I}}\cup\{\epsilon\}$, are defined by
\begin{align*}
	[R]_{\sigma_1,\sigma_2}=
	\begin{cases}
		\mathcal{A}((\sigma_1,\sigma_2)), & \text{if} \quad(\sigma_1,\sigma_2)\in R, \\
		\infty, & \text{if} \quad(\sigma_1,\sigma_2)\notin R.
	\end{cases}
\end{align*}

Given an ERM $[R]$ w.r.t. $\Sigma_{\mathcal{I}}\cup\{\epsilon\}$, the error function with cost, denoted by $\mathcal{R}:\Sigma_{\mathcal{I}}^*\rightarrow 2^{(\Sigma_{\mathcal{I}}\times \mathbb{Z}_{\geq 0})^*}$, is inductively defined as follows:
\begin{itemize}

	\item $\mathcal{R}(\epsilon)=\{(\sigma,[R]_{\epsilon,\sigma})|\exists\sigma\in\Sigma_{\mathcal{I}}:[R]_{\epsilon,\sigma}\neq\infty\}$;
	\item $\forall \sigma\in\Sigma_{\mathcal{I}}:\mathcal{R}(\sigma)=\{(\sigma',[R]_{\sigma,\sigma'})|\exists\sigma'\in\Sigma_{\mathcal{I}}\cup\{\epsilon\}:[R]_{\sigma,\sigma'}$ $\neq\infty\}$;
	\item $\forall\sigma\omega\in\Sigma_{\mathcal{I}}^*:\mathcal{R}(\sigma\omega)=\mathcal{R}(\epsilon)^*\mathcal{R}(\sigma)\mathcal{R}(\epsilon)^*\mathcal{R}(\omega)$.
\end{itemize}

As usual, $\mathcal{R}(\epsilon)^*$ denotes the set of all finite concatenations of the set $\mathcal{R}(\epsilon)$. 
Note that, the last item implies that there exist insertion actions before and after an observation. 
Given a sequence $\omega\in\Sigma_{\mathcal{I}}^*$, $\mathcal{R}(\omega)$ represents the set of all possible erroneous sequences w.r.t. the sequence $\omega$, each of which has the form $(\sigma_1,c_1)\dots(\sigma_i,c_i)\dots(\sigma_n,c_n)$ where $\sigma_i\in\Sigma_{\mathcal{I}}\cup\{\epsilon\}$ and $c_i\in \mathbb{Z}_{\geq 0}$. 
Given a sequence $\hat{\omega}_r=(\sigma_1,c_1)\dots(\sigma_i,c_i)\dots(\sigma_n,c_n)\in\mathcal{R}(\omega)$, we use\footnote{Despite the potential presence of $\epsilon$ in $\sigma_1\dots\sigma_i\dots\sigma_n$, it is irrelevant in string comparisons since $s\cdot\epsilon=\epsilon\cdot s=s$ for any string $s\in\Sigma^*$. The same logic will be applied in subsequent sections.} $\omega_r=\sigma_1\dots\sigma_i\dots\sigma_n$ and $c_{\omega_r}=\Sigma^n_{i=1}c_i$ to denote the corresponding sequence of events and its total cost, respectively. 
Thus, the set of erroneous sequences of events with costs, denoted by $\overline{\mathcal{R}}(\omega)$, is defined as $\overline{\mathcal{R}}(\omega)=\{(\omega_r,c_{\omega_r})|\hat{\omega}_r\in\mathcal{R}(\omega)\}$. 

Under the assumption that there exists an upper bound on the cost of errors, denoted by a non-negative integer $c_u\in\mathbb{Z}_{\geq 0}$, upon any sequence $\omega$, $\overline{\mathcal{R}}(\omega)$ can be re-defined by $\overline{\mathcal{R}}^{c_u}(\omega)=\{(\omega_r,c_{\omega_r})| \hat{\omega}_r\in\mathcal{R}(\omega)\wedge c_{\omega_r}\leq c_u\}$.
We use $[R]^{c_u}$ to denote an ERM $[R]$ with the cost constraint $c_u$ and $\mathbf{C_u}=\{0,1,\dots,c_u\}$ is the corresponding set of costs. 
We will omit the superscript ``$c_u$'' in the description of entries of $[R]^{c_u}$ when it is clear from the context.
Note that for all $\sigma\in\Sigma_{\mathcal{I}}, [R]_{\sigma,\sigma}=0$, i.e., there should be an error-less action for each event in $\Sigma_{\mathcal{I}}$.

\begin{example}\label{E1-withERM}
	Given a set of events $\Sigma_{\mathcal{I}}=\{\alpha_{12}, \beta_{13}, \sigma_2, \gamma_3\}$ (the subscripts of events match the indices of OSs that observe the specific events), consider the following ERM $[R]^{2}$ w.r.t. $\Sigma_{\mathcal{I}}\cup\{\epsilon\}$ (the superscript indicates the cost constraint). 
\vspace{-0.2cm}
	\[
	\scriptsize
	[R]^2 = 
	\bordermatrix{ & \epsilon & \alpha_{12} & \beta_{13} & \sigma_2 & \gamma_3\cr
		\epsilon&  0     & 1                  & \infty           & \infty        & \infty          \cr
		\alpha_{12}&  \infty     & 0                 & \infty          & 1                & \infty          \cr
		\beta_{13}&  1            & \infty           & 0                 & \infty         & \infty           \cr
		\sigma_2&  \infty     & \infty           & \infty          & 0                & \infty           \cr
		\gamma_3&  \infty     &  \infty          & 1                 & \infty          & 0     }           \qquad
	\]
\vspace{-0.4cm}

The ERM presents four error actions, each of which has one unit of cost, and five error-less actions (since $|\Sigma_{\mathcal{I}}\cup\{\epsilon\}|=5$). 
Note that $[R]_{\epsilon,\alpha_{12}}=1$ indicates that symbol $\alpha_{12}$ can be inserted at any position with a cost of one unit; $[R]_{\beta_{13},\epsilon}=1$ indicates that symbol $\beta_{13}$ can be deleted with a cost of one unit. 
The same principle applies to other error actions in $[R]^{2}$. 
Given a sequence $\omega=\alpha_{12}\sigma_2\beta_{13}\beta_{13}\gamma_3$, the set of possible erroneous sequences with costs, based on the error function, is  $\overline{\mathcal{R}}^2(\omega)=\{(\omega, 0), (\alpha_{12}\sigma_2\beta_{13}\gamma_3, 1),(\alpha_{12}\sigma_2\beta_{13}\beta_{13}\beta_{13}, 1),\allowbreak(\alpha_{12}\sigma_2\beta_{13}\beta_{13}\gamma_3\alpha_{12}\alpha_{12}, 2),\dots \}$.\hfill\rule{1ex}{1ex}
\end{example}

\section{State Estimation Under Global Errors}

The global error is used to describe a scenario where the occurrence of an error influences the observations of all OSs which can observe the erroneous symbol.
A typical assumption in this case is that the error action $(\sigma_1,\sigma_2)\in R$ only affects the OSs whose sets of observable events contain both $\sigma_1$ and $\sigma_2$. 
Here, however, we forego this assumption in the case of global error. 
This means we allow any error action $(\sigma_1,\sigma_2)\in R$ to occur, for two reasons: first, it is a more comprehensive and general setting; second, the methods presented in this section do not have any restriction on addressing this general setting. 
These discussions lead to the following definition.

Given an ERM $[R]^{c_u}$, suppose that a sequence of events $t$ is executed by a system. Let $(\omega_r,c_{\omega_r})\in\overline{\mathcal{R}}^{c_u}(P_{\mathcal{I}}(t))$ be the corrupted sequence with its associated cost. The erroneous SI-state under global error (E$_g$SI-state), received at the coordinator, is $\tau_g=(\tau^{(1)}_g,\dots, \tau^{(m)}_g)$, where $\tau^{(i)}_g=P_i(\omega_{r})$.
We use $\omega_{r}$ to denote the possible erroneous sequence w.r.t. $P_{\mathcal{I}}(t)$, which indicates that  errors occurred before being recorded at each OS and $\tau^{(i)}_g=P_i(\omega_{r})$ denotes the corresponding PO-sequences. With this definition, the error-tolerant state estimation problem under consideration in this section is formulated below.

\begin{figure*}[htbp]
	\centering
	\includegraphics[scale=0.9]{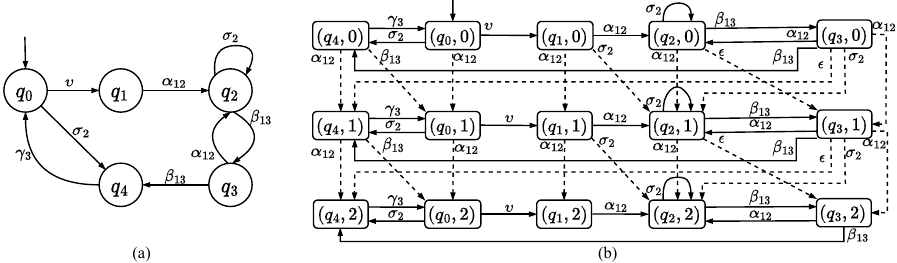}
	\caption{(a) NFA model $G$ where $\Sigma_1=\{\alpha_{12},\beta_{13}\}$, $\Sigma_2=\{\alpha_{12},\sigma_2\}$, and $\Sigma_3=\{\beta_{13}, \gamma_3\}$ discussed in Example \ref{E1-withERM} and (b) modified system $G_g$ w.r.t. the ERM $[R]^2$ (dotted lines used to indicate that these transitions are the results of error actions).}
	\label{system model}
\end{figure*}

\begin{problem}
	(DO-based state estimation under bounded global error (DO-E$_g$SE)) Given system $G=(Q,\Sigma,\delta,Q_0)$, bounded global errors may happen w.r.t. an ERM $[R]^{c_u}$: following a string $t\in\Sigma^*$, the E$_g$SI-state received at the coordinator is $\tau_g=(\tau^{(1)}_g,\dots, \tau^{(m)}_g)$, where $\tau^{(i)}_g=P_i(\omega_r)$ with $\omega_r\in\overline{\mathcal{R}}^{c_u}(P_{\mathcal{I}}(t))$. 
    If $\omega_r$ results in a synchronization, the coordinator needs to compute the set of error-tolerant state estimates
	\begin{multline*}
\hspace{-0.43cm}
		\mathcal{E}^{c_u}_g(\tau_g,Q_0)=\{(q,c)\in Q\times \mathbf{C_u}|\exists q_0\in Q_0,\exists u\in L(G,q_0):q\in\\
		\delta(q_0,u)\land(\exists(\omega_r,c)\in\overline{\mathcal{R}}^{c_u}(P_{\mathcal{I}}(u)),\forall i\in\mathcal{I}:P_i(\omega_r)=\tau_g^{(i)})\}.
	\end{multline*}Note that one can also focus on calculating the set of least cost error-tolerant state estimates
	$	\mathcal{E}^{c_u,m}_g(\tau_g,Q_0)=\{(q,c_m)\in\mathcal{E}^{c_u}_g(\tau_g,Q_0)|\nexists(q,c)\in\mathcal{E}^{c_u}_g(\tau_g,Q_0): c<c_m\},$
	but we do not explicitly discuss this case in the remainder of this paper.
\end{problem}

\subsection{DO-E$_g$SE from the Perspective of System Modification}
The basic idea is to modify a system by encoding all possible error actions based on an ERM $[R]^{c_u}$. Then, the DO-E$_g$SE could be performed using the standard method presented in \cite{SunHadjicostisLi2023} for the case of DO-based state estimation. 
Given an NFA $G=(Q,\Sigma,\delta, Q_0)$, the cost-constrained globally modified system w.r.t. an ERM $[R]^{c_u}$, is a four-tuple NFA, denoted by $G_g=(Q^g_{c_u}, \Sigma\cup\{\epsilon\}, \delta_g, Q^g_{0, c_u})$, where $Q^g_{c_u}\subseteq Q\times\mathbf{C_u}$ is the set of states, $\Sigma\cup\{\epsilon\}$ is the set of events, $Q^g_{0, c_u}=Q\times\{0\}\subseteq Q^g_{c_u}$ is the set of initial states, and $\delta_g:Q^g_{c_u}\times(\Sigma\cup\{\epsilon\})\rightarrow 2^{Q^g_{c_u}}$ is the state transition function, defined as follows: for any $(q,c)\in Q^g_{c_u}$, $\sigma\in\Sigma\cup\{\epsilon\}$, one has $\delta_g((q,c),\sigma)=\Delta_{g_1}\cup \Delta_{g_2}\cup \Delta_{g_3}$, where
\begin{align*}
	\Delta_{g_1} &=
	\begin{cases}
		\delta(q,\sigma)\times \{c\}, & \text{if}\enspace\sigma\in\Sigma\setminus\Sigma_{\mathcal{I}}\land\delta(q,\sigma)! \\
		\emptyset, & \text{otherwise},
	\end{cases} \\
	\Delta_{g_2} &=
	\begin{cases}
		\delta(q,\sigma')\times \{c+[R]_{\sigma',\sigma}\}, &
		\text{if}\enspace\exists\sigma'\in\Sigma_{\mathcal{I}}: c+[R]_{\sigma',\sigma}\\&\leq c_u\land 
\delta(q,\sigma')! \\
		\emptyset, & \text{otherwise},
	\end{cases} \\
	\Delta_{g_3} &=
	\begin{cases}
		\{(q,c+[R]_{\epsilon,\sigma})\}, & \text{if}\enspace \sigma\neq\epsilon\land c+[R]_{\epsilon,\sigma}\leq c_u, \\
		\emptyset, & \text{otherwise}.
	\end{cases}
\end{align*}

The inequality predicates $c+[R]_{\sigma',\sigma}\leq c_u$ and $c+[R]_{\epsilon,\sigma}\leq c_u$ in the above function indirectly capture the necessity for  $\sigma\in R(\sigma')$ and $\sigma\in R(\epsilon)$, respectively. The same logic will be applied throughout this work. The cost component of each state represents the total cost of error actions upon a possible string from an initial state to that particular state.
If we focus on the cost-constrained globally modified system $G_g$ which is known to be in the states $Q_0\times\{0\}$, the state estimation of $G_g$ after the coordinator receives $\tau_g$ is given by 
\begin{multline*}
	\mathcal{E}_{G_g}(\tau_g,Q_0\times\{0\})=\{(q,c)|\exists (q_0,0)\in Q_0\times\{0\},\exists u\in \\L(G_g, (q_0,0)):P_{\mathcal{I}}(u)\in \mathcal{TO}(\tau_g) \wedge (q,c)\in\delta_g((q,0), u)\},
\end{multline*} where $\mathcal{TO}(\tau_g)$ is the set of TO-sequences based on the PO-sequences of $\tau_g$ in $G_g$ (see (1) in Section \Romannum{2}.B).

\begin{lemma}\label{G-A-Lemma}
	Consider a sequence of events $t$ occurring in system $G$ such that the E$_g$SI-state is $\tau_g$ w.r.t. an ERM $[R]^{c_u}$. Given its modified system $G_g$, the DO-E$_g$SE after the coordinator receives $\tau_g$ is given by 
	\begin{align*}
	\mathcal{E}^{c_u}_g(\tau_g,Q_0)=\mathcal{E}_{G_g}(\tau_g,Q_0\times\{0\}).
	\end{align*}
\end{lemma}

The proof of the above lemma, as well as those of the other conclusions, can be found in the Appendix.
Lemma~\ref{G-A-Lemma} implies that the problem of DO-E$_g$SE is reduced to the problem of DO-based state estimation in the modified system, where the E$_g$SI-state $\tau_g$ is viewed as a normal SI-state. In \cite{SunHadjicostisLi2023}, state estimation under the DO-based protocol is achieved through the construction of the synchronizer via a breadth-first search based on the notion of S-builder. 
Here, we use E$_g$-synchronizer $\mathcal{S}_g(\tau_{g},Q_0\times\{0\})=(T,\Sigma_s,T_0,T_e,h_s, c_g,C_g)$ to denote such a synchronizer of the modified system $G_g$ w.r.t. an E$_g$SI-state $\tau_g$ and a set of states $Q_0\times\{0\}$, where $c_g:T\rightarrow 2^{Q^g_{c_u}}$ is the state estimation function, $C_g$ summarizes this mapping for each state of the S-builder, and $T_e$ is the marked state. We initialize by setting $T_0=\tau_g$ and $c_g(T_0)=\operatorname{UR}(Q_0\times\{0\})$. For simplicity, we use $\mathcal{S}_g$ instead of $\mathcal{S}_g(\cdot)$ when it is clear in the context. Readers can refer to \cite{SunHadjicostisLi2023} for more details. 

\begin{theorem}\label{theorem-global-1}
	Given an E$_g$SI-state $\tau_g$ w.r.t. an ERM $[R]^{c_u}$, let $\mathcal{S}_g( \tau_g, Q_0\times\{0\}) = ( T, \Sigma_s, T_0, T_e, h_s, c_g, C_g )$ be the corresponding synchronizer of the modified system $G_g$. It holds $\mathcal{E}^{c_u}_g(\tau_g, Q_0)=c_g(T_e)$, where $c_g(T_e)$ is the set of state estimates assigned to $T_e$.
\end{theorem}

\begin{figure}[tbp]
	\centering
	\includegraphics[scale=0.9]{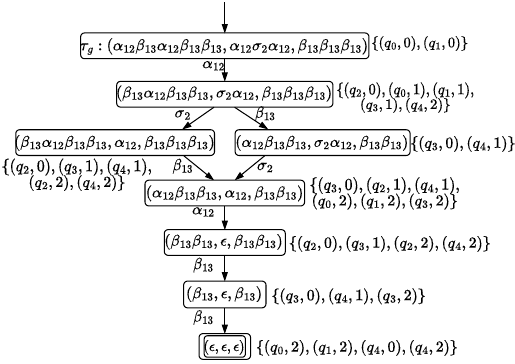}
	\caption{The synchronizer of the modified system $G_g$ w.r.t. the given $\tau_{g}$ and $Q_0=\{q_0\}$, in which  $c_g(\tau_g)=\operatorname{UR}(Q_0\times\{0\})=\{(q_0,0),(q_1,0)\}$. The corresponding set of state estimates is displayed next to each SI-state.}
	\label{global-1-s}
\end{figure}

\begin{example}\label{E-G1}
Consider the system $G$ shown in Fig. \ref{system model}(a) with $\mathcal{I}=\{1,2,3\}$, where $\Sigma_1=\{\alpha_{12},\beta_{13}\}$, $\Sigma_2=\{\alpha_{12},\sigma_2\}$, and $\Sigma_3=\{\beta_{13}, \gamma_3\}$. Let us consider the ERM $[R]^2$ shown in Example~\ref{E1-withERM}. The corresponding modified system $G_g$ w.r.t. $[R]^2$ for $G$ is shown in Fig. \ref{system model}(b).

Suppose that string $t=\upsilon\alpha_{12}\sigma_2\beta_{13}\beta_{13}\gamma_3\in L(G)$ (which is unknown to the coordinator) occurs in the system originating from state $q_0$ and ending at state $q_0$. If, at this time instant, a synchronization is initiated such that the coordinator receives $\tau_g=(\alpha_{12}\beta_{13}\alpha_{12}\beta_{13}\beta_{13},\alpha_{12}\sigma_2\alpha_{12}, \beta_{13}\beta_{13}\beta_{13})$, we know that there exists $(\omega_r, c_{\omega_r})\in\overline{\mathcal{R}}^{c_u}(P_{\mathcal{I}}(t))$ such that $\tau_g^{(i)}=P_i(\omega_r)$. By Theorem \ref{theorem-global-1}, Algorithm 1 in \cite{SunHadjicostisLi2023} is applied to construct the corresponding synchronizer of the modified system $G_g$, as shown in Fig. \ref{global-1-s}, where the state estimates assigned to each SI-state are positioned alongside them. Then, we obtain that DO-E$_g$SE is $\mathcal{E}^{c_u}_g(\tau_g,Q_0)=c_g((\epsilon,\epsilon,\epsilon))=\{(q_0,2),(q_1,2),(q_4,0),(q_4,2)\}$. \hfill\rule{1ex}{1ex}
\end{example}

\subsection{DO-E$_g$SE from the Perspective of S-builder Modification}

If the error model (ERM $[R]^{c_u}$) changes (e.g., if the costs are time-varying), the coordinator needs to re-construct the modified system accordingly to utilize the previous approach. This leads to increased computational complexity and additional difficulty. 
To this end, the method proposed in this subsection concentrates on the inference of possible matching TO-sequences (called global-error-tolerant TO-sequences), which may be tampered such that the E$_g$SI-states is $\tau_g$. 
Based on the possible matching TO-sequences, error-tolerant state estimates can be directly obtained using the original system without requiring any additional modification. 
Here, we first define the global-error-tolerant TO-sequences.
\begin{definition}\label{global-def-TO}
	(Global-error-tolerant TO-sequences (GETO-sequences)) Suppose that a sequence of events $t$ occurs in a system $G$, such that the E$_g$SI-state is $\tau_g=(\tau^{(1)}_g,\dots, \tau^{(m)}_g)$ w.r.t. an ERM $[R]^{c_u}$. The set of GETO-sequences upon $\tau_g$, denoted by $\mathcal{TO}^{c_u}_g(\tau_g)$, is defined as a subset of sequences of events in $\Sigma_{\mathcal{I}}^*$ with associated costs, i.e.,
	\begin{multline*}
		\mathcal{TO}^{c_u}_g(\tau_g)=\{(\omega,c)\in\Sigma_{\mathcal{I}}^*\times\mathbf{C_u}|\exists\omega_r\in\Sigma^*_{\mathcal{I}}:\\\omega_r\in\mathcal{TO}(\tau_g)\land(\omega_r,c)\in\overline{\mathcal{R}}^{c_u}(\omega)\}.
	\end{multline*}
\end{definition}

\begin{lemma}\label{global-lemma-state}
	Consider a sequence of events $t$ occurring in system $G$ such that the E$_g$SI-state is $\tau_g$ w.r.t. an ERM $[R]^{c_u}$. The DO-E$_g$SE after the coordinator receives $\tau_g$ is given by 
	\begin{multline*}
		\mathcal{E}^{c_u}_g(\tau_g,Q_0)=\{(q,c)|\exists q_0\in Q_0, \exists u\in L(G,q_0): \\
		(P_{\mathcal{I}}(u),c)\in\mathcal{TO}^{c_u}_g(\tau_g)\land q\in\delta(q_0,u)\}.
	\end{multline*}
\end{lemma}

The proof of the above lemma follows directly from the definitions of $\mathcal{E}^{c_u}_g(\tau_g,Q_0)$ and $\mathcal{TO}^{c_u}_g(\tau_g)$, and is therefore omitted here.
Intuitively, for any $(\omega, c)$ in $\mathcal{TO}^{c_u}_g(\tau_g)$, $\omega$ represents a sequence of observations in the system that may be tampered with, such that the  E$_g$SI-state is $\tau_g$ and $c$ is the corresponding cost.
If all GETO-sequences can be inferred, the problem of DO-E$_g$SE can be performed directly on the original system. 
Based on the notion of S-builder, a global-error-tolerant-sequence-builder is defined below whose marked language can be used to infer GETO-sequences for a given E$_g$SI-state $\tau_g$.

\begin{definition}\label{E$_g$TS-builder}
	(Global-error-tolerant-sequence-builder (E$_g$TS-builder)) Given an E$_g$SI-state $\tau_g$ w.r.t. an ERM $[R]^{c_u}$, an E$_g$TS-builder is a five-tuple transition system $\widetilde{B}_g=(\widetilde{T}_g,\Sigma^p,\widetilde{T}_{g,0},\widetilde{T}_{g,e},\widetilde{h}_g)$, where
	\begin{itemize}
		\item $\widetilde{T}_g\subseteq T\times\mathbf{C_u}$ is the set of E$_g$SI-states associated with costs;
		\item $\Sigma^p\subseteq(\Sigma_{\mathcal{I}}\cup\{\epsilon\})\times(\Sigma_{\mathcal{I}}\cup\{\epsilon\})\setminus(\{\epsilon\}\times\{\epsilon\})$ is the set of pairs of events, where $\sigma^p_{[0]}$ and $\sigma^p_{[1]}$ denote the first and second components of an event $\sigma^p\in\Sigma^p$, respectively;
		\item $\widetilde{T}_{g,0}=(\tau_g,0)\in \widetilde{T}_g$ is the initial state with zero cost;
		\item $\widetilde{T}_{g,e}\subseteq \{T_e\}\times\mathbf{C_u}$ is the set of marked and ending states associated with possibly different costs;
		\item $\widetilde{h}_g:\widetilde{T}_g\times \Sigma^p\rightarrow \widetilde{T}_g$ is the deterministic event release transition function, defined as follows: for any $(\tau,c)=(\tau^{(1)},\ldots,\tau^{(m)},c), (\tau',c')=(\tau'^{(1)},\ldots,\tau'^{(m)},c')\in \widetilde{T}_g$, $\sigma^p=(\sigma^p_{[0]},\sigma^p_{[1]})\in\Sigma^p$, it holds
		\begin{multline*}
\hspace{-0.2cm}
			\widetilde{h}_g((\tau,c),\sigma^p)=(\tau',c')\Rightarrow \text{H1}\lor \text{H2} \\
\hspace{-0.2cm}
			\shoveleft{\text{H1}\Leftrightarrow \sigma^p_{[1]}=\epsilon\land\tau=\tau'\land c'=c+[R]_{\sigma^p_{[0]},\sigma^p_{[1]}}\leq c_u}\\
\hspace{-0.2cm}
			\shoveleft{\text{H2}\Leftrightarrow \sigma^p_{[1]}\neq\epsilon\land h(\tau,\sigma^p_{[1]})=\tau'\land} c'=c+[R]_{\sigma^p_{[0]},\sigma^p_{[1]}}\leq c_u.
		\end{multline*}
	\end{itemize}
\end{definition}

The domain of function $\widetilde{h}_g$ can be extended to $\widetilde{T}_g\times(\Sigma^p)^*$ in the standard recursive manner: $\widetilde{h}_g((\tau_{g1},c),\sigma^pt^p)=\widetilde{h}_g(\widetilde{h}_g((\tau_{g1},c),\sigma^p),t^p)$ for $(\tau_{g1},c)\in\widetilde{T}_g$, $\sigma^pt^p\in(\Sigma^p)^*$. 

Since an error action occurs globally to a specific event, the mechanism used by the E$_g$TS-builder is identical to that of the S-builder. 
In other words, the transition function $\widetilde{h}_g$ constructs the state space of the E$_g$TS-builder recursively by releasing the events in $\tau_g$. 
The distinction lies in the assumption that each released event results from either an error-less or error action.
Consequently, we define the set of transition symbols as a set of event pairs, $\Sigma^p$.
Given $\sigma^p\in\Sigma^p$, $\sigma^p_{[1]}$ denotes the released event in a specific E$_g$SI-state, while $\sigma^p_{[0]}$ represents the possible event in the sequence of original observations. 
Therefore, (H1) holds when $\sigma^p_{[0]}$ is assumed to be deleted from the original sequence. 
As a result, no event is removed from $\tau$, and the only difference lies in the cost component, i.e., $\tau=\tau'$ and $c'=c+[R]_{\sigma^p_{[0]},\sigma^p_{[1]}}$.
Similarly, (H2) holds in other situations where $\sigma^p_{[1]}$ can be released in $\tau$, i.e., $h(\tau,\sigma^p_{[1]})=\tau'$.
In this case, the event $\sigma^p_{[1]}$ can be the result of an error-less action (where $\sigma^p_{[0]}=\sigma^p_{[1]}$), an error action of replacement (where $\sigma^p_{[0]}\neq\epsilon$ and $\sigma^p_{[0]}\neq\sigma^p_{[1]}$), or an error action of insertion (where $\sigma^p_{[0]}=\epsilon$).
Further, we release the event $\sigma^p_{[1]}$ in $\tau$ and the transition is $(\sigma^p_{[0]},\sigma^p_{[1]})$. 
Note that the cost associated with each E$_g$SI-state refers to the total cost of error actions upon the path from the initial state to this particular state. 
Due to error actions of deletion, there may be multiple possible ending states associated with different costs. 
The following result follows directly from Definition \ref{E$_g$TS-builder}.

\begin{lemma}\label{lemma-Monotonicity property}
	(Monotonicity property) Let $(\tau,c)$ and $(\tau',c')$ be two E$_g$SI-states with costs in an E$_g$TS-builder. If there exists $\sigma^p\in\Sigma^p$ such that $(\tau',c')=\widetilde{h}_g((\tau,c),\sigma^p)$, then it holds
	\begin{align}
		(\forall i\in\mathcal{I}:|\tau'^{(i)}|\leq|\tau^{(i)}|)\land c'\geq c.
	\end{align}
\end{lemma}

For any synchronization, there exists $k_i\in\mathbb{N}$, such that $|\tau^{(i)}|\leq k_i$, i.e., the memory capacity of each OS is finite. When paired with the cost constraint $c_u\in\mathbb{Z}^+$ and the monotonicity property, it can be inferred from Definition~\ref{E$_g$TS-builder} that the state space of an E$_g$TS-builder $\widetilde{B}_g$ is finite. Letting $t^p=\sigma^p_1\dots\sigma^p_n\in L(\widetilde{B}_g)$, we use $t^p[0]$ and $t^p[1]$ to denote the sequences formed by the first and second elements of the pairs in $t^p$, respectively. In this case, we have the following lemma.

\begin{lemma}\label{global-GETO}
	Given an E$_g$SI-state $\tau_g$ w.r.t. an ERM $[R]^{c_u}$, and its corresponding E$_g$TS-builder $\widetilde{B}_g$, it holds:
	\begin{enumerate}
		\item $\forall t^p\in L_m(\widetilde{B}_g):((T_e,c)=\widetilde{h}_g(\widetilde{T}_{g,0},t^p)\Rightarrow(t^p[1],c)\in\overline{\mathcal{R}}^{c_u}(t^p[0]))$;
		\item $\mathcal{TO}^{c_u}_g(\tau_g)=\{(t^p[0],c)|\exists t^p\in L_m(\widetilde{B}_g):(T_e,c)=\widetilde{h}_g(\widetilde{T}_{g,0},t^p)\}$.
	\end{enumerate}
\end{lemma}



According to Lemma \ref{global-GETO}.2), the GETO-sequences for any $\tau_g$ can be obtained directly from its corresponding E$_g$TS-builder.
Given this, an intuitive approach to perform DO-E$_g$SE is to compute the product of the E$_g$TS-builder and the system observer w.r.t. the set of observable events $\Sigma_{\mathcal{I}}$. 
Though useful, this requires the full state spaces of both structures, which leads to high computational complexity.
Instead, we propose that DO-E$_g$SE is carried out concurrently with the construction of the E$_g$TS-builder.  
This approach, referred to as \textit{estimation-by-release}, was also utilized in \cite{SunHadjicostisLi2023}. 
The E$_g$TS-builder aims to infer the possible sequences of observations through the \textit{release} procedure.
Our goal is to incorporate the \textit{estimation} procedure into the construction of the E$_g$TS-builder.
The basic concept applied to the error scenario is outlined as follows.

We first assign the latest state estimates to state $(\tau_g,0)$.
While constructing the E$_g$TS-builder from $(\tau_g,0)$, we simultaneously calculate and assign to each E$_g$SI-state a set of state estimates representing the system's possible states following the observation sequences from $(\tau_g,0)$ to this particular E$_g$SI-state, i.e., $t^p[0]$ for any $t^p$ in the corresponding E$_g$TS-builder.
Each transition $((\tau,c),\sigma^p,(\tau',c'))\in \widetilde{h}_g$ signifies the release of $\sigma^p_{[1]}$ at $(\tau,c)$ and  new state estimates are assigned to $(\tau',c')$ by updating the set of states assigned to $(\tau,c)$ through the observation $\sigma^p_{[0]}$ (which implies that $((\tau,c),\sigma^p,(\tau',c'))$ is not feasible if $\sigma^p_{[0]}$ is not defined at any system state in the state estimates assigned to $(\tau,c)$). 
Since for any E$_g$SI-state $\tau$ with cost $c$, there may exist multiple transitions originating from different states and ending at $(\tau,c)$, the state estimation of $(\tau,c)$ involves different updated state estimates and assignments. 
Therefore, the state estimation of a specific E$_g$SI-state is only performed if no other transition ends at it in subsequent release processes. 
Thus, we conclude the following rules.

\textit{Release rule}: An E$_g$SI-state is eligible for the release procedure if and only if it has been assigned a state estimate.

\textit{Estimation rule}: An E$_g$SI-state with cost can undergo the state estimation procedure if and only if no further release transitions terminate at it in the subsequent release processes.


Based upon the above rules, Algorithm \ref{E$_g$T-synchronizer} is proposed to construct the structure called an E$_g$T-synchronizer, where the main body is the sub-structure of an E$_g$TS-builder, each state of which is assigned a set of system state estimates. 
The E$_g$T-synchronizer is denoted by $\mathcal{\widetilde{S}}_g(\tau_g,Q_0)=(\widetilde{T}_g,\Sigma^p,\widetilde{T}_{g,0},\widetilde{T}_{g,e},\widetilde{h}_{g,s},\widetilde{c}_g,\widetilde{C}_g)$, where $(\widetilde{T}_g,\Sigma^p,\widetilde{T}_{g,0},\widetilde{T}_{g,e},\widetilde{h}_{g,s})$ is a sub-structure of the E$_g$TS-builder, $ \widetilde{c}_g:\widetilde{T}_g\rightarrow 2^{Q}$ is the state-estimation function, $\widetilde{C}_g$ summarizes this mapping for each state in $\mathcal{\widetilde{S}}_g$, and $\widetilde{T}_{g,e}$ is the set of ending and marked states.
In order to satisfy the two rules, we propose a two-level hierarchical release strategy, involving breadth-first search (BFS) over sequence components and depth-first search (DFS) over cost components from the initial state $(\tau_g,0)$. Hereafter, we use ``layer'' to refer to the depth-level in BFS.

\begin{algorithm}
	\caption{Construction of an E$_g$T-synchronizer}\label{E$_g$T-synchronizer}
	\begin{algorithmic}[1]
		\Require System $G$, set of states $Q_0$, E$_g$SI-state $\tau_g$, ERM $[R]^{c_u}$.
		\Ensure $\mathcal{\widetilde{S}}_g(\tau_g,Q_0)=(\widetilde{T}_g,\Sigma^p,\widetilde{T}_{g,0},\widetilde{T}_{g,e},\widetilde{h}_{g,s},\widetilde{c}_g,\widetilde{C}_g)$.
		\State $\widetilde{T}_{g,0}=(\tau_g,0)$, $\widetilde{T}_g=\{\widetilde{T}_{g,0}\}$, $\widetilde{c}_g(\widetilde{T}_{g,0})=\operatorname{UR}(Q_0)$, $\widetilde{C}_g=\{\widetilde{c}_g(\widetilde{T}_{g,0})\}$, $\widetilde{T}_{g,e}=\emptyset$, add tag ``estimate'' to $(\tau_g,0)$;
		\While{$\widetilde{T}_{g,e}==\emptyset$}
		\State \hspace{-0.3cm}$\widetilde{T}_{\text{tag}}$ is the set of states with tags ``estimate'';
		\State  \hspace{-0.3cm}\textit{Estimation}($\widetilde{T}_{\text{tag}}$);
		\State \hspace{-0.3cm}$\widetilde{T}_{g,e}=\{(T_e,c)\in \widetilde{T}_g\}$;
		\EndWhile
		\State \textbf{procedure} \textit{Estimation}($\widetilde{T}_{1}$)
		\For{$\{(\tau,c)\in \widetilde{T}_1|\nexists(\tau,c')\in \widetilde{T}_1:c'<c\}$}
		\State \hspace{-0.3cm}\textit{Esti\&DFS}($(\tau,c)$);
		\EndFor
		\State \textbf{procedure} \textit{Esti\&DFS}($(\tau,c)$)
		\If{$(\tau,c)\neq(\tau_g,0)$}
		\State \hspace{-0.3cm} $\widetilde{c}_g((\tau,c))=\bigcup_{((\tau',c'),\sigma^p,(\tau,c))\in\widetilde{h}_{g,s}}\operatorname{UR}(\operatorname{R}_{\sigma^p_{[0]}}(\widetilde{c}_g(\tau',c')))$;
		\State \hspace{-0.3cm}$\widetilde{C}_g=\widetilde{C}_g\cup\{\widetilde{c}_g((\tau,c))\}$;
		\EndIf
		\State Remove tag ``estimate'' of $(\tau,c)$ and \textit{Release}($(\tau,c)$);
		\If{$\exists(\tau,c_1)$: $\widetilde{c}_g((\tau,c_1))=\emptyset$, such that $\nexists(\tau,c_2)$: $c_2<c_1 \wedge \widetilde{c}_g((\tau,c_2))=\emptyset$}
		\State \hspace{-0.3cm}\textit{Esti\&DFS}($(\tau,c_1)$);
		\EndIf
		\State \textbf{procedure} \textit{Release}($(\tau,c)$)
		\For{$\sigma^p\in\{\sigma^p|\widetilde{h}_g((\tau,c),\sigma^p)!\land\operatorname{R}_{\sigma^p_{[0]}}(\widetilde{c}_g((\tau,c)))\neq\emptyset\}$}
		\State \hspace{-0.3cm}add $\widetilde{h}_g((\tau,c),\sigma^p)$ and $((\tau,c),\sigma,\widetilde{h}_g((\tau,c),\sigma^p))$ to $\widetilde{T}_g$ and $\widetilde{h}_{g,s}$, respectively; add tag ``estimate'' to $\widetilde{h}_g((\tau,c),\sigma^p)$;
		\EndFor
	\end{algorithmic}
\end{algorithm}


\begin{figure*}[htbp]
	\centering
	\includegraphics[scale=1.2]{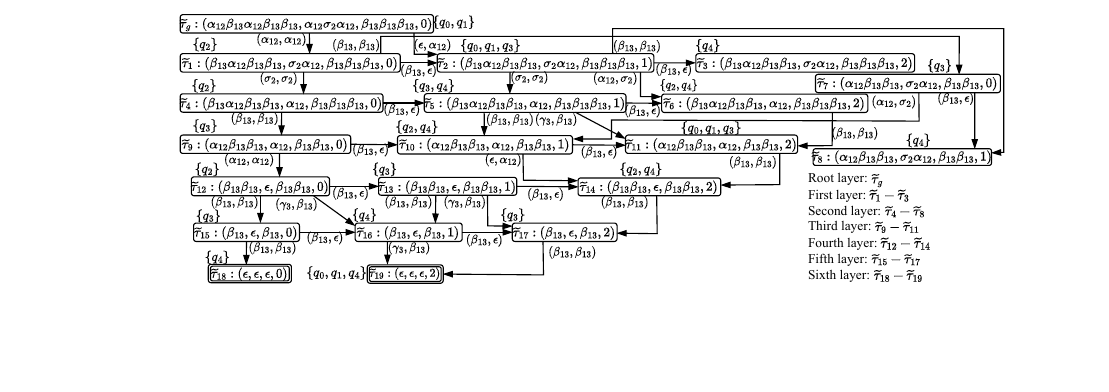}
	\caption{The E$_g$T-synchronizer of the system $G$ w.r.t. the given $\tau_{g}$ and $Q_0$. The state estimate is displayed next to each state, where $\widetilde{T}_{g,0}=\widetilde{\tau}_g$ is the initial state with $\widetilde{c}_g(\widetilde{\tau}_g)=\operatorname{UR}(Q_0)=\{q_0,q_1\}$ and $\widetilde{T}_{g,e}=\{\widetilde{\tau}_{18}, \widetilde{\tau}_{19}\}$ is the set of marked and ending states. The states in each layer are listed to the side.}
	\label{global-2-s}
\end{figure*}

Starting from the initial state $(\tau_g,0)$,  viewed as the root node, the release rule is easily fulfilled by performing the release procedure immediately after the state estimation for a specific E$_g$SI-state with cost. 
Due to the monotonicity property, there are two distinct types of release transitions represented by $((\tau, c), \sigma^p, (\tau', c'))\in h_g$. 
The first type occurs when $\tau=\tau'$ and $c'>c$, i.e., (H1) holds. 
This type suggests that event $\sigma^p[0]$ is assumed to be deleted in the system observations. 
In this case, $(\tau', c')$ is listed at the same layer as $(\tau, c)$. 
The second type occurs when there exists $i\in\mathcal{I}$,  $\tau'^{(i)}<\tau^{(i)}$ and $c'\geq c$, i.e., (H2) holds. 
This type encompasses other situations not covered by the first type. 
In this case, $(\tau', c')$ is listed at the next layer after $(\tau, c)$. 
By examining these two types of release transitions, the approach guarantees that an E$_g$SI-state in the current layer cannot release an event that would terminate at a state in a previous layer. 
Additionally, it ensures that distinct E$_g$SI-states within the same layer do not involve transitions to each other.

Combining the aforementioned release procedure rules, the assignment of state estimates is carried out alongside the event release procedure by using depth-first search (DFS) to explore a specific E$_g$SI-state with different costs.
Within each layer, we prioritize the state of the form $(\tau,c)$, where $c$ is the least cost among all states with the same $\tau$ (lines 6--8). 
By utilizing the DFS starting from the state $(\tau, c)$ (lines 14--15), state estimation is performed and assigned to the state $(\tau, c)$ (line 11).
Following the state estimation, the release process is carried out using the \textit{Release} procedure (lines 16--18).
Within this procedure, $\widetilde{h}_g((\tau,c),\sigma^p)!$ and $\operatorname{R}_{\sigma^p[0]}(\widetilde{c}_g(\tau, c))\neq\emptyset$ indicate that not only is $\sigma^p$ defined at $(\tau,c)$ in the E$_g$TS-builder, but also the first component of $\sigma^p$, i.e., $\sigma^p[0]$, is defined at (at least) one state in $\widetilde{c}_g(\tau, c)$ in the original system.
The reason behind employing a depth-first approach is to handle the error action of deletion, i.e., condition (H1) in Definition \ref{E$_g$TS-builder}, allowing an E$_g$SI-state to be associated with different costs within the same layer.
This ensures that the state estimation and release operations are conducted in a comprehensive manner, capturing all possible states at each level of the breadth-first search.

\begin{theorem}\label{theorem-global-2}
	Given an E$_g$SI-state $\tau_g$ w.r.t. an ERM $[R]^{c_u}$, let $\mathcal{\widetilde{S}}_g(\tau_g, Q_0) = (\widetilde{T}_g, \Sigma^p, \widetilde{T}_{g,0}, \widetilde{T}_{g,e}, \allowbreak\widetilde{h}_{g,s}, \widetilde{c}_g, \widetilde{C}_g)$ be the corresponding E$_g$T-synchronizer. It holds $\mathcal{E}^{c_u}_g(\tau_g,Q_0) = \{(q,c)|\exists(T_e,c)\in\widetilde{T}_{g,e}: q\in \widetilde{c}_g((T_e,c))\}$.
\end{theorem}

\begin{example}
Let us still consider system $G$, string $t$, and E$_g$SI-state $\tau_g$ shown in Example \ref{E-G1}. 
Algorithm \ref{E$_g$T-synchronizer} is used to construct the corresponding E$_g$T-synchronizer $\mathcal{\widetilde{S}}_g(\tau_g,\{q_0\})$, as shown in Fig. \ref{global-2-s}. 
For the initial state $\widetilde{\tau}_g$, the transitions $(\epsilon,\alpha_{12})$ and $(\alpha_{12},\alpha_{12})$ are admissible, where $\alpha_{12}$ could be the original observation or the inserted events, since $\operatorname{R}_{\epsilon}(\{q_0,q_1\})\neq\emptyset$ and  $\operatorname{R}_{\alpha_{12}}(\{q_0,q_1\})\neq\emptyset$.
However, transition $(\beta_{13},\epsilon)$, an error action of deletion, is not possible here even though $\widetilde{h}_g(\widetilde{\tau}_g,(\beta_{13},\epsilon))!$ due to  $\operatorname{R}_{\beta_{13}}(\{q_0,q_1\})=\emptyset$. 
In this case, $\widetilde{\tau}_g$ is the only state in this layer. 
In the next layer, we start procedure \textit{Esti\&DFS} for $\widetilde{\tau}_1$ which has the least cost of E$_g$SI-state $(\beta_{13}\alpha_{12}\beta_{13}\beta_{13},\sigma_2\alpha_{12}, \beta_{13}\beta_{13}\beta_{13})$. 
After the execution of line 13, the state estimates of $\widetilde{\tau}_1$ are computed as $\widetilde{c}_g(\widetilde{\tau}_1)=\operatorname{UR}(\operatorname{R}_{\alpha_{12}}(\widetilde{c}_g(\widetilde{\tau}_g)))=\operatorname{UR}(\operatorname{R}_{\alpha_{12}}(\{q_0,q_1\})=\{q_2\}$. 
After state estimation, we immediately start procedure \textit{Release} from $\widetilde{\tau}_1$, where the transitions $(\sigma_2,\sigma_2)$, $(\beta_{13},\beta_{13})$, and $(\beta_{13},\epsilon)$ end at $\widetilde{\tau}_4$, $\widetilde{\tau}_7$, and $\widetilde{\tau}_2$, respectively. 
Then, procedure \textit{Esti\&DFS} is again executed upon $\widetilde{\tau}_2$ which has the least cost among the states sharing the same E$_g$SI-state that have not yet been assigned state estimates. 
In this case, $\widetilde{c}_g(\widetilde{\tau}_2)=\operatorname{UR}(\operatorname{R}_{\beta_{13}}(\widetilde{c}_g(\widetilde{\tau}_1)))\cup\operatorname{UR}(\operatorname{R}_{\epsilon}(\widetilde{c}_g(\widetilde{\tau}_g)))=\{q_0,q_1,q_3\}$. 
The procedures \textit{Release} and \textit{Esti\&DFS} are subsequently executed upon $\widetilde{\tau}_2$ and $\widetilde{\tau}_3$, respectively.  
We then have state $\widetilde{\tau}_3$ and its assigned state estimates $\{4\}$. 
After these steps, this layer contains three states, i.e., $\widetilde{\tau}_1$, $\widetilde{\tau}_2$, and $\widetilde{\tau}_3$.
The remainder of this E$_g$T-synchronizer can be constructed in a similar way. 
Note that $\widetilde{T}_{g,e}=\{\widetilde{\tau}_{18}, \widetilde{\tau}_{19}\}$ is the set of ending states and $\mathcal{E}^{c_u}_g(\tau_g,Q_0)=\{(q_4,0)\}\cup(\{q_0,q_1,q_4\}\times\{2\})=\{(q_0,2),(q_1,2),(q_4,0),(q_4,2)\}$. 
These estimates are the same as those computed in Example~\ref{E-G1}.\hfill\rule{1ex}{1ex}
\end{example}

\section{State Estimation Under Local Errors}

A local error is a commonly occurring phenomenon in systems and can be viewed as a more general scenario compared with a global error. 
A local error differs in the sense that a single error action impacts the information reported by one of the OSs. 
Consequently, the sequences of events received at the coordinator may be subject to diverse error actions.

We assume the existence of an error relation $R_i$ for each OS, denoted by $\{R_i\}_{i\in\mathcal{I}}$, where $\mathcal{I}$ is the index set. 
The definition of each of $\{R_i\}_{i\in\mathcal{I}}$ is similar to the one presented in Section \Romannum{3}. 
For any $i\in\mathcal{I}$, $R_i$ is defined on $\Sigma_i$, so that the ERM $[R_i]$ and the corresponding error function with cost is of the form $\mathcal{R}_i:\Sigma^*_i\rightarrow 2^{(\Sigma_i\times \mathbb{Z}_{\geq 0})^*}$. 
We also assume there is an upper bound cost $c_u$ that represents the cumulative cost of erroneous sequences reported by all OSs during a synchronization step. 
Hereafter, $\{[R_i]\}^{c_u}_{i\in\mathcal{I}}$ is referred to as the set of local ERMs and the superscript  $c_u$ is the corresponding total cost constraint.

Given $\{[R_i]\}^{c_u}_{i\in\mathcal{I}}$, suppose that a sequence of events $t$ is executed by the system. 
The erroneous SI-state under local error (E$_l$SI-state), received at the coordinator, is $\tau_l=(\tau_l^{(1)},\dots,\tau_l^{(m)})$, where for any  $i\in\mathcal{I}$, $(\tau_l^{(i)},c_i)\in\overline{\mathcal{R}}_i(P_i(t))$ and $\Sigma^m_{i=1}c_i\leq c_u$. 
Here, the notation $(\tau^{(i)}_l,c_i)\in\overline{\mathcal{R}}_i(P_i(t))$ indicates that the sequence recorded by each OS is tampered separately and $\Sigma^m_{i=1}c_i\leq c_u$ indicates that the sum of costs over all PO-sequences is below $c_u$. 
With this definition, the local error-tolerant state estimation problem under consideration in this section is formulated as follows.

\begin{problem}
	(DO-based state estimation under bounded local error (DO-E$_l$SE)) Given system $G=(Q,\Sigma,\delta,Q_0)$, bounded local errors may happen w.r.t. $\{[R_i]\}^{c_u}_{i\in\mathcal{I}}$ during one synchronization. Following a string $t\in\Sigma^*$, which occurs in the system and results in a synchronization, the E$_l$SI-state received at the coordinator is $\tau_l=(\tau^{(1)}_l,\dots, \tau^{(m)}_l)$, where $(\tau_l^{(i)},c_i)\in\overline{\mathcal{R}}_i(P_i(t))$ and $\Sigma^m_{i=1}c_i\leq c_u$. The coordinator needs to compute the set of error-tolerant state estimates
	\begin{multline*}
		\mathcal{E}^{c_u}_l(\tau_l,Q_0)=\{(q,c)\in Q\times \mathbf{C_u}|
		\exists q_0\in Q_0,\exists u\in L(G,q_0):\\q\in\delta(q_0,u) \land
		( \forall i\in\mathcal{I},\exists c_i\in\mathbf{C_u}:
		(\tau_l^{(i)},c_i)\in\overline{\mathcal{R}}_i(P_i(u))\land\\ c=\Sigma^m_{i=1}c_i\leq c_u)\}.
	\end{multline*}
\end{problem}

\subsection{DO-E$_l$SE from the Perspective of System Modification}
Each time an event occurs in the system, at most $m$ OSs could observe and record this observation. 
The original system model does not  incorporate this information, but this can be easily accomplished by extending the events of transitions into $m$-tuples.
Given an NFA $G=(Q,\Sigma,\delta,Q_0)$, we can incorporate the observation information by introducing an observation automaton $G_o=(Q,\Sigma^m,\delta_o, Q_0)$ for the DO-based protocol, where $Q$ is the set of system states, $Q_0$ is the set of initial states, $\Sigma^m=\{\sigma^m=(\sigma^{(1)},\dots,\sigma^{(m)})|\sigma\in\Sigma,\forall i\in\mathcal{I}:\sigma^{(i)}=P_i(\sigma)\}$ is the set of $m$-tuple events, and $\delta_o:Q\times\Sigma^m\rightarrow 2^Q$ is the transition function, where for all $q,q'\in Q$, $(\exists\sigma\in\Sigma:(q,\sigma,q')\in\delta)\Leftrightarrow(\exists\sigma^m\in\Sigma^m:(q,\sigma^m,q')\in\delta_o)$. 
Obviously, for all $\sigma\in\Sigma\setminus\Sigma_{\mathcal{I}}$, $\sigma^m=(\epsilon,\dots,\epsilon)\in\Sigma^m$.

The only distinction between $G$ and $G_o$ lies in the expansion of transitions of $G$ into $m$-tuples in $G_o$. 
The $m$ components of transitions in $G_o$ can be viewed as the information transmitted by different communication channels between sensors and OSs. This allows for a more detailed description of how the occurrences of events in a system affect the recorded sequence at each OS. 
To improve the performance of $G_o$ so that it can capture the influence of local errors, we incorporate the ERMs into the transition relations of $G_o$. 
Given an observation automaton $G_o=(Q,\Sigma^m,\delta_o, Q_0)$, the cost-constrained locally modified system w.r.t. $\{[R_i]\}^{c_u}_{i\in\mathcal{I}}$, is a four-tuple NFA, denoted by $G_l=(Q^l_{c_u},\Sigma^m_l,\delta_l,Q^l_{0,c_u})$, where
\begin{itemize}
	\item $Q^l_{c_u}\subseteq Q\times\mathbf{C_u}$ is the set of states;
	\item $\Sigma^m_l=\Sigma^m\cup\Sigma^m_{l_{d,r}}\cup\Sigma^m_{l_{in}}\subseteq(\Sigma_1\cup\{\epsilon\})\times\dots\times(\Sigma_m\cup\{\epsilon\})$ is the set of $m$-tuple events with:
	\begin{itemize}
		\item $\Sigma^m_{l_{d,r}}=\{(e^{(1)},\dots,e^{(m)})|\exists\sigma^m\in\Sigma^m,\forall i\in\mathcal{I}:(\sigma^{(i)}=\epsilon\Rightarrow e^{(i)}=\epsilon)\land(\sigma^{(i)}\neq\epsilon\Rightarrow e^{(i)}\in R_i(\sigma^{(i)}))\}$;
		\item $\Sigma^m_{l_{in}}=\{(e^{(1)},\dots,e^{(m)})|\exists i\in\mathcal{I}:e^{(i)}\in R_i(\epsilon)\setminus\{\epsilon\}\land(\forall j\in\mathcal{I}\setminus\{i\}:e^{(j)}=\epsilon)\}$.
	\end{itemize}
	\item $Q^l_{0,c_u}=Q\times\{0\}\subseteq Q^l_{c_u}$ is the set of initial states;
	\item $\delta_l: Q^l_{c_u}\times\Sigma^m_l\rightarrow 2^{Q^l_{c_u}}$ is the transition function, defined for any $(q,c)\in Q^l_{c_u}$, 
$e^m=(e^{(1)},\dots,e^{(m)})\in\Sigma^m_l$ as $\delta_l((q,c),e^m)=\Delta_{l_1}\cup \Delta_{l_2}\cup \Delta_{l_3}$ where
	\begin{align*}
		\Delta_{l_1} &=
		\begin{cases}
			\delta_o(q,e^m)\times\{c\}, & \text{if} \quad \mathcal{CD}1,\\
			\emptyset, & \text{otherwise},
		\end{cases} \\
		\Delta_{l_2} &=
		\begin{cases}
			\delta_o(q,\sigma^m)\times\\\{c+\Sigma_{i\in\mathcal{I}\land\sigma^{(i)}\neq\epsilon}[R_i]_{\sigma^{(i)},e^{(i)}}\},&\text{if}\:\mathcal{CD}2,\\
			\emptyset, &\text{otherwise},
		\end{cases} \\
		\Delta_{l_3} &=
		\begin{cases}
			\{(q,c+[R_i]_{\epsilon,e^{(i)}})\}, &\text{if} \quad \mathcal{CD}3,\\
			\emptyset, & \text{otherwise},
		\end{cases}
	\end{align*}
	with $\mathcal{CD}1$, $\mathcal{CD}2$, and $\mathcal{CD}3$ defined by:
	\begin{itemize}
		\setlength{\itemindent}{-0.8cm}
		\item[] $\mathcal{CD}1: e^m\in\Sigma^m\wedge \delta_o(q,e^m)!$;
		\item[] $\mathcal{CD}2: e^m\in\Sigma^m_{l_{d,r}}\land\sigma^m\in\Sigma^m\land(\exists j\in\mathcal{I}:\sigma^{(j)}=\epsilon\Rightarrow e^{(j)}=\epsilon)
\land (c+\Sigma_{i\in\mathcal{I}\land\sigma^{(i)}\neq\epsilon}[R_i]_{\sigma^{(i)},e^{(i)}}\leq c_u)
\land\delta_o(q,\sigma^m)!$;
		\item[] $\mathcal{CD}3:e^m\in\Sigma^m_{l_{in}}\land (\exists i\in\mathcal{I},\exists e^{(i)}\neq \epsilon:c+[R]_{\epsilon,e^{(i)}}\leq c_u)$.
	\end{itemize}
\end{itemize}


In the process of constructing $G_l$, we use $e^m\in\Sigma^m_l$ to represent the possibly tampered information that could be recorded by the OSs for each $\sigma^m\in\Sigma^m$ (note that $\Sigma^m\subseteq\Sigma^m_l$).  
The set $\Sigma^m_{l_{in}}$ consists of $m$-tuple transitions that are considered to be erroneously inserted in the PO-sequences. 
For any $e^m\in\Sigma^m_{l_{in}}$, only one component in $e^m$ is not $\epsilon$ (we explain this in Remark \ref{RE-OneInsertion}). 
The set $\Sigma^m_{l_{d,r}}$ consists of $m$-tuple transitions whose components represent the outcomes of erroneous replacements and/or deletions in the PO-sequences. 
The transition function $\delta_l$ describes the possible state evolution based on the transitions in $\Sigma^m_l$. $\Delta_{l_1}$ is the set of next states if the transition $e^m$ is assumed to be error-free,  $\Delta_{l_2}$ represents the set of next states if there exists a transition $\sigma^m \in \Sigma^m$ where the non-$\epsilon$ components (i.e., the components that are not $\epsilon$) can be replaced or deleted to obtain the transition $e^m$, and $\Delta_{l_3}$ is the set of next states after the error actions of insertions have occurred. 
Thus, in any state of $\Delta_{l_3}$, only the cost component is changed.
\begin{remark}\label{RE-OneInsertion}
The function $\delta_l$ simulates various scenarios when error (-less) actions in $\{[R_i]\}^{c_u}_{i\in\mathcal{I}}$ occur in specific $m$-tuple transitions. 
It is natural to question why there is only one non-$\epsilon$ component in $e^m\in\Sigma^m_{l_{in}}$, indicating that only one error action of insertion is considered at a time. 
Similarly, one may wonder why there are no error actions of insertions in $e^m\in\Sigma^m_{l_{d,r}}$. 
These design choices are made to avoid redundancy and simplify the analysis. 
Consider states $(q,c),(q,c'')\in Q^l_{c_u}$, for which there exists an event $e^m=(\epsilon,\dots,\epsilon,\sigma^{(i)},\epsilon,\dots,\sigma^{(j)},\epsilon\dots,\epsilon)$ indicating that symbols $\sigma^{(i)}$ and $\sigma^{(j)}$ are inserted in the recorded PO-sequences of $O_i$ and $O_j$, respectively, such that $((q,c),e^m,(q,c''))\in\delta_l$ with $c''=c+[R_i]_{\epsilon,\sigma^{(i)}}+[R_j]_{\epsilon,\sigma^{(j)}}$. 
Then, we can simplify this procedure by using two separate transitions: $e^m_1=(\epsilon,\dots,\sigma^{(i)},\dots\epsilon)$ and $e^m_2=(\epsilon,\dots,\sigma^{(j)},\dots\epsilon)$. 
This approach allows us to simulate the situation in two steps: $((q,c),e^m_1,(q,c'))\in\delta_l$ with $c'=c+[R_i]_{\epsilon,\sigma^{(i)}}$ and $((q,c'),e^m_2,(q,c''))\in\delta_l$ with $c''=c'+[R_j]_{\epsilon,\sigma^{(j)}}$. 
Effectively, the need for the existence of $e^m$ is eliminated, as we can always find $e^m_1$ and $e^m_2$ that serve the same purpose. 
The same reasoning can be applied in the absence of error actions of insertions in $\Sigma^m_{l_{d,r}}$.
\end{remark}

Let $t=\sigma_1\dots\sigma_n\in L(G,q)$. 
The corresponding string in $G_o$ is $t^m=(P_1(\sigma_1),\dots,P_m(\sigma_1))\dots(P_1(\sigma_n),\dots,$ $P_m(\sigma_n))$. 
We define $l_i(t^m)$, $i\in\mathcal{I}$, to be the restriction of $t^m$ to each of its components, i.e., $l_i(t^m)=P_i(\sigma_1)\dots P_i(\sigma_n)=\sigma_1^{(i)}\dots\sigma_n^{(i)}$. 
Given a modified system $G_l$ which is known to be in the set of states $Q_0\times\{0\}$, the state estimation of $G_l$ after the coordinator receives $\tau_l$ is defined as $\mathcal{E}_{G_l}(\tau_l,Q_0\times\{0\})=\{(q,c)|\exists (q_0,0)\in Q^l_{0,c_u}, \exists s^m\in L(G_l,(q_0,0)),\forall i\in\mathcal{I}:l_i(s^m)=\tau^{(i)}_l\wedge (q,c)\in\delta_l((q_0,0),s^m)\}$. 

\begin{lemma}\label{local-A-state}
	Given an E$_l$SI-state $\tau_l=(\tau^{(1)}_l,\dots,\tau^{(m)}_l)$, it holds
	\begin{align*}
		\mathcal{E}^{c_u}_l(\tau_l,Q_0)=\mathcal{E}_{G_l}(\tau_l,Q_0\times\{0\}).
	\end{align*}
\end{lemma}

\begin{definition}
	(Multi-S-builder (MS-builder)) Given an E$_l$SI-state $\tau_l$ w.r.t. $\{[R_i]\}^{c_u}_{i\in\mathcal{I}}$, an MS-builder is a five-tuple transition system $B^m=(T,\Sigma^m_s,T_0,T_e,h^m)$, where
	\begin{itemize}
		\item $T\subseteq \Sigma_1^*\times\dots\times\Sigma_m^*$ is the set of E$_l$SI-states;
		\item $\Sigma^m_s=\Sigma^m_l\setminus\{(\epsilon,\dots,\epsilon)\}$ is the set of $m$-tuple events;
		\item $T_0\in T$ is the initial state with $T_0=\tau_l$;
		\item $T_e\in T$ is the marked ending state;
		\item $h^m:T\times \Sigma^m_s\rightarrow T$ is the event release transition function, which is defined as follows: for any $\tau,\tau'\in T$, $\tau\neq\tau'$, and  $e^m=(e^{(1)},\dots,e^{(m)})\in\Sigma^m_s$, it holds
		\begin{align*}
			h^m(\tau,e^m)=\tau'\Rightarrow \forall i\in\mathcal{I}:\tau'^{(i)}=\tau^{(i)}/e^{(i)}.
		\end{align*}
	\end{itemize}
\end{definition}

The marked language of the MS-builder captures all possible strings that can occur at some states of the $G_l$ upon E$_l$SI-state $\tau_l$. 
Next, we show how to apply the \textit{estimation-by-release} approach described in Section \Romannum{4}.B to the state estimation problem here. 
The monotonicity property in an MS-builder is as follows: for any release transition $(\tau,e^m,\tau')\in h^m$, for all $i\in\mathcal{I}$, $|\tau'^{(i)}|\leq|\tau^{(i)}|$. 
We aim to construct the corresponding E$_l$-synchronizer $\mathcal{S}_l$ w.r.t. the specific $\tau_l$ and set of system states. 
There is one notable difference between $h^m$ and $h_s$ of the E$_g$-synchronizer $\mathcal{S}_g$: in an $\mathcal{S}_g$, different E$_g$SI-states within the same layer, determined through the BFS, release transitions that terminate at E$_g$SI-states in the same subsequent layer. 
In contrast, release transitions from one E$_l$SI-state may terminate at E$_l$SI-states in different subsequent layers. 
This difference arises due to the nature of the function $h^m$. 
Therefore, we define the counting function $N: T \rightarrow \mathbb{Z}_{\geq0}$, where for all $\tau \in T$, $N(\tau) = \sum_{i=1}^{m} |\tau^{(i)}|$. 
This function is used to label the order of the estimation procedure for different $\tau$ based on their total number of events. 
According to the monotonicity property of the states in an MS-builder, the E$_l$SI-states with the largest number of events and without assigning the system states satisfy the \textit{estimation rule}. 
Therefore, in Algorithm~\ref{E$_l$-synchronizer}, the E$_l$-synchronizer $\mathcal{S}_l(\tau_l,Q_0\times\{0\})=(T,\Sigma^m_l,T_0,T_e,h^m_s,c_l,C_l)$ is constructed using the \textit{estimation-by-release} approach, taking into account this difference, where $c_l:T\rightarrow 2^{Q^l_{c_u}}$ is the state estimation function and $c_l(T_0)=\operatorname{UR}(Q_0\times\{0\})$.

\begin{algorithm}
	\caption{Construction of an E$_l$-synchronizer}\label{E$_l$-synchronizer}
	\begin{algorithmic}[1]
		\Require Modified system $G_l$, state set $Q_0\times\{0\}$, E$_l$SI-state $\tau_l$.
 		\Ensure $\mathcal{S}_l(\tau_l,Q_0\times\{0\})=(T,\Sigma^m_l,T_0,T_e,h^m_s,c_l,C_l).$
		\State $T_0=\tau_l$, $T=\{T_0\}$, $c_l(T_0)=\operatorname{UR}(Q_0\times\{0\})$, $C_l=\{c_l(T_0)\}$, Release Level $RL=\emptyset$;
		\State \textit{Release}($\tau_l$);
		\While{$RL\neq \emptyset$}
		\State \textit{Estimation}($\{\tau|N(\tau)=\max(RL)\}$);
		\State $RL=RL\setminus \{\max(RL)\}$;
		\EndWhile
		\State \textbf{procedure} \textit{Estimation}($T_1$)
		\For{$\tau\in T_1$}
		\State $c_l(\tau)=\bigcup_{(\tau',e^m,\tau)\in h^m_s}\operatorname{UR}(\operatorname{R}_{e^m}(c_l(\tau')))$;
		\State $C_l=C_l\cup\{c_l(\tau)\}$;
		\State \textit{Release}($\tau$);
		\EndFor
		\State \textbf{procedure} \textit{Release}($\tau$)
		\For{ $e^m\in\Sigma^m_l$ s.t. $h^m(\tau,e^m)!\land \operatorname{R}_{e^m}(c_l(\tau))\neq\emptyset$}
		\State add $h^m(\tau,e^m)$ and $(\tau,e^m,h^m(\tau,e^m))$ to $T$ and $h^m_s$, respectively, $RL=RL\cup\{N(h^m(\tau,e^m))\}$;
		\EndFor
	\end{algorithmic}
\end{algorithm}

\begin{remark}
The DO-E$_l$SE problem differs from its counterpart in the case of global error, where the relevant TO-sequences are inferred. 
In particular, the first events of all given $m$ sequences are processed at the same time as long as these $m$ events collectively form an $m$-tuple event, as defined in the relevant states of the modified system. 
Given that there exists an ERM for each channel from the system to a specific OS, an event that is simultaneously observable by two OSs can now potentially be recorded as two distinct events. 
Therefore, in this subsection, we incorporate these various possibilities into the modified system, wherein the MS-builder functions solely as a systematic model designed to address the sequence. 
\end{remark}
\begin{figure*}[htbp]
	\centering
	\subfigure[]{\includegraphics[scale=0.8]{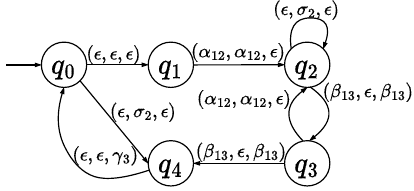}} 
	\subfigure[]{\includegraphics[scale=0.97]{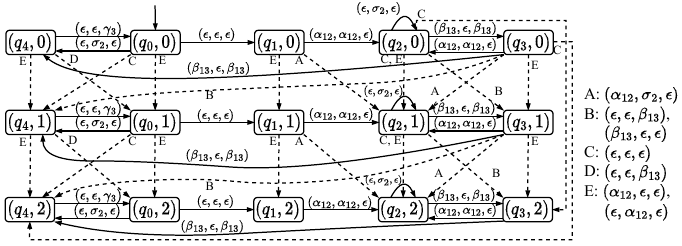}} 
	\caption{(a) The observation automaton $G_o$ of $G$ in Fig. \ref{E1-withERM}(a) and (b) the modified system $G_l$ w.r.t. the ERM $\{[R_i]\}^2_{i\in\{1,2,3\}}$ in Example \ref{EX-local-1} (dotted lines are used to indicate that these transitions are the result of error actions; to keep the diagram concise, the transitions in the modified system are represented by labels A-E, shown on the right).}
	\label{local-system}
\end{figure*}

\begin{figure}[htbp]
	\centering
	\includegraphics[scale=0.87]{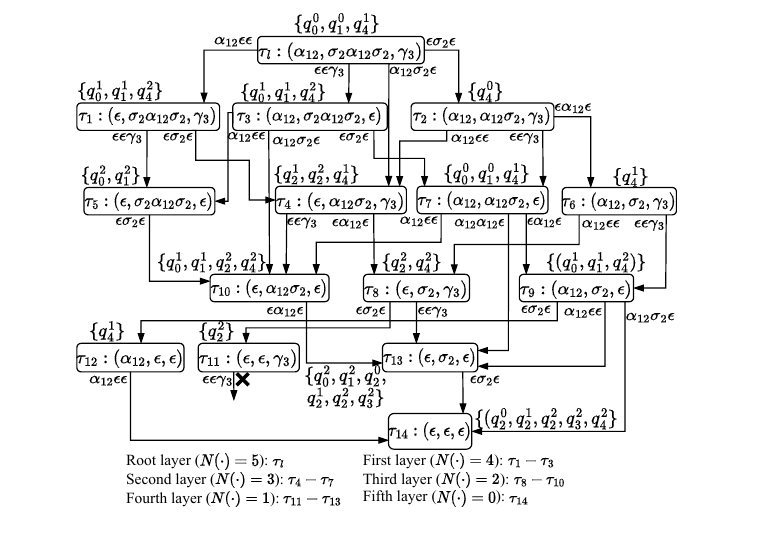}
	\caption{The E$_l$-synchronizer of the modified system $G_l$ w.r.t. the given $\tau_{l}$ and $\operatorname{UR}(Q_0\times\{0\})=\{(q_0,0),(q_1,0),(q_4,1)\}$. For the sake of brevity, in this diagram, we write the state with cost $(q,c)$ in the form of  $q^c$. The corresponding set of state estimates is displayed next to each E$_l$SI-state and states in each layer are listed at the bottom of the figure.}
	\label{local-1-s}
\end{figure}

\begin{theorem}\label{local-A-theorem}
	Given an E$_l$SI-state $\tau_l$ w.r.t. $\{[R_i]\}^{c_u}_{i\in\mathcal{I}}$, let $\mathcal{S}_l(\tau_l,Q_0\times\{0\})=(T,\Sigma^m_l,T_0,T_e,h^m_s,c_l,C_l)$ be the corresponding E$_l$-synchronizer of the modified system $G_l$. It holds $\mathcal{E}^{c_u}_l(\tau_l,Q_0)=c_l(T_e)$.
\end{theorem}
The proof of the above theorem follows from Lemma~\ref{local-A-state} and Theorem~\ref{theorem-global-1} and is therefore omitted here.

\begin{example}\label{EX-local-1}
Consider again the system in Fig.~\ref{E1-withERM}(a). 
Its corresponding observation automaton $G_o$ is shown in Fig. \ref{local-system}(a). 
Consider the following ERMs $\{[R_i]\}^2_{i\in\{1,2,3\}}$:
	\noindent 
	\begin{minipage}{0.33\linewidth}
		\centering 
		$[R_1]$
		\[
		\scriptsize
		\bordermatrix{  & \epsilon  & \alpha_{12} & \beta_{13}        \cr
			\epsilon  &  0     & 1                   & \infty                  \cr
			\alpha_{12}&  \infty     & 0                   & \infty                   \cr      
			\beta_{13} &  1            & \infty            & 0                          \cr }   
		\]
	\end{minipage}
	\begin{minipage}{0.33\linewidth}
		\centering
		$[R_2]$
		\[
		\scriptsize
		\bordermatrix{  & \epsilon  & \alpha_{12} & \sigma_2        \cr
			\epsilon          &  0      & 1                  & \infty                  \cr
			\alpha_{12}    &  \infty      & 0                  & 1                         \cr      
			\sigma_2       &  1             & \infty            & 0                          \cr }     
		\]
	\end{minipage}\begin{minipage}{0.33\linewidth}
		\centering
		$[R_3]$
		\[
		\scriptsize
		\bordermatrix{  & \epsilon  & \beta_{13} & \gamma_3        \cr
			\epsilon          &  0      & \infty          & \infty                  \cr
			\beta_{13}     &  1              & 0                & \infty                  \cr      
			\gamma_3     &  \infty       & 1                & 0                          \cr }           
		\]
	\end{minipage}  

	\vspace{0.15cm}

\noindent where $[R_1]$, $[R_2]$, and $[R_3]$ represent the ERMs for three OSs, respectively, with an upper bound cost of two in a single synchronization step. 
The corresponding modified system is depicted in Fig.~\ref{local-system}(b).  
Suppose that the coordinator receives $\tau_l=(\alpha_{12},\sigma_2\alpha_{12}\sigma_2,\gamma_3)$. First, we obtain the unobservable reach of the set $Q_0\times\{0\}$ and assign it to $\tau_l$, i.e., $\operatorname{UR}(Q_0\times\{0\})=\{(q_0,0),(q_1,0),(q_4,1)\}$. The corresponding E$_l$-synchronizer is portrayed in Fig. \ref{local-1-s}. For the sake of brevity, in this diagram, we represent the transition $(x_1,x_2,x_3)$ as $x_1x_2x_3$ and the state with cost $(q,c)$ as $q^c$, respectively. 
Using the counting function $N$, we could fulfill the \textit{estimation rule} when constructing the synchronizer. 
Despite the direct connection between $\tau_l$ and $\tau_4$ via $\alpha_{12}\sigma_2\epsilon$, the estimation procedure first operates on the set of states $\{\tau_1,\tau_2,\tau_3\}$ in the first layer before $\tau_4$, following the \textit{Release} procedure at state $\tau_l$, since some transitions to $\tau_4$ may still be unreleased.
Based on the discussion in Remark \ref{RE-OneInsertion}, suppose a hypothetical transition $((q_4,1),(\alpha_{12},\epsilon,\gamma_3),(q_0,2))$, where $\alpha_{12}$ is inserted in the original transition $(\epsilon,\epsilon,\gamma_3)$ defined at state $(q_4,1)$. 
In the E$_l$-synchronizer shown in Fig. \ref{local-1-s}, this transition can be interpreted as having the same effect in the estimation process as transitions $(\tau_l,(\alpha_{12},\epsilon,\epsilon),\tau_1),(\tau_1,(\epsilon,\epsilon,\gamma_3),\tau_5)\in h^m_s$.\hfill\rule{1ex}{1ex}

\end{example}

\subsection{DO-E$_l$SE from the Perspective of S-builder Modification}

In this subsection, we alternatively explore the possible TO-sequences of the system without requiring any modifications, which leads to the following definition.
\begin{definition}\label{local-def-TO}
	(Local-error-tolerant TO-sequences (LETO-sequences)) Suppose that a sequence of events $t$ occurs in $G$, such that the E$_l$SI-state is $\tau_l=(\tau^{(1)}_l,\dots, \tau^{(m)}_l)$ w.r.t. $\{[R_i]\}^{c_u}_{i\in\mathcal{I}}$. The set of LETO-sequences upon $\tau_l$, denoted by $\mathcal{TO}^{c_u}_l(\tau_l)$, is defined as a subset of sequences of events in $\Sigma_{\mathcal{I}}^*$ with associated costs, i.e., 
	\begin{multline*}
		\mathcal{TO}^{c_u}_l(\tau_l)=\{(\omega,c)\in\Sigma_{\mathcal{I}}^*\times\mathbf{C_u}|
		\exists \omega\in\Sigma^*_{\mathcal{I}},
		\forall i\in\mathcal{I},\\
		\exists c_i\in\mathbf{C_u}:
		(\tau_l^{(i)},c_i)\in\overline{\mathcal{R}}_i(P_i(\omega))\land
		c=\Sigma^m_{i=1}c_i\leq c_u\}.
	\end{multline*}
\end{definition}

\begin{lemma}\label{local-lemma-state}
	Consider a sequence of events $t$ occurring in system $G$ such that the E$_l$SI-state is $\tau_l$ w.r.t. $\{[R_i]\}^{c_u}_{i\in\mathcal{I}}$. The DO-E$_l$SE after the coordinator receives $\tau_l$ is
	\begin{multline*}
		\mathcal{E}^{c_u}_l(\tau_l,Q_0)=\{(q,c)|\exists q_0\in Q_0, \exists u\in L(G,q_0): \\
		(P_{\mathcal{I}}(u),c)\in\mathcal{TO}^{c_u}_l(\tau_l)\land q\in\delta(q_0,u)\}.
	\end{multline*}
\end{lemma}

The proof of the above lemma follows directly from the definitions of $\mathcal{E}^{c_u}_l(\tau_l,Q_0)$ and $\mathcal{TO}^{c_u}_l(\tau_g)$, and is therefore omitted.
Similar to Definition \ref{global-def-TO} and Lemma \ref{global-lemma-state}, Definition \ref{local-def-TO} captures the set of possible sequences of original events with their associated costs, denoted as $\mathcal{TO}^{c_u}_{l}(\tau_l)$, and Lemma \ref{local-lemma-state} indicates that \mbox{DO-E$_l$SE} can be directly obtained from $\mathcal{TO}^{c_u}_{l}(\tau_l)$.
Unlike the standard S-builder and E$_g$TS-builder,  the complexity of inferring $\mathcal{TO}^{c_u}_{l}(\tau_l)$ increases since different error actions can occur across different PO-sequences.
Here, we introduce the notion of \textit{release list}, which is defined upon PO-sequences provided by OSs. 
For any $i\in\mathcal{I}$, given $\alpha\omega\in\Sigma^*_i$, we denote by $\Phi_i(\alpha\omega)=E_{i,D}(\alpha\omega)\dot{\cup}E_{i,R}(\alpha\omega)\dot{\cup} E_{i,I}(\alpha\omega)\dot{\cup}\{(\alpha,\alpha)\}$ the \textit{release list} of pairs of possible original and received events upon the string $\alpha\omega$, where $E_{i,D}(\alpha\omega)=\{(\sigma,\epsilon)|\sigma\neq\epsilon,\epsilon\in R_i(\sigma)\}$\footnote{For any  $\phi\in E_{i,D}(\cdot)$, both components of $\phi$ are unrelated to ``$\cdot$'', which is a natural outcome since we assume the first component of $\phi$ is deleted in the original sequence. It is necessary to maintain this assumption at any position in the PO-sequences while ensuring that the cost constraints are satisfied.},
$E_{i,R}(\alpha\omega)=\{(\sigma,\alpha)|\sigma\neq\epsilon,\alpha\neq\epsilon,\sigma\neq\alpha,\alpha\in R_i(\sigma)\}$, and $E_{i,I}(\alpha\omega)=\{(\epsilon,\alpha)|\alpha\neq\epsilon,\alpha\in R_i(\epsilon)\}$ (i.e., the release list is defined upon the first event of $\alpha\omega$ with the convention that the first event appears on the left). 
The first component of each pair in $E_{i,D}(\cdot)$ ($E_{i,R}(\cdot)$ or $E_{i,I}(\cdot)$, respectively) represents the possible original event generated by the system, which is assumed to be deleted (replaced or inserted, respectively). 
The second component indicates the received event in the PO-sequence at this specific position.  

\begin{definition}\label{E$_l$TS-builder}
	(Local-error-tolerant-sequence-builder (E$_l$TS\allowbreak-builder)) Given an E$_l$SI-state $\tau_l$ w.r.t. $\{[R_i]\}^{c_u}_{i\in\mathcal{I}}$, an E$_l$TS-builder is a five-tuple transition system $\widetilde{B}_l=(\widetilde{T}_l,\Sigma^{pm},\widetilde{T}_{l,0},\widetilde{T}_{l,e},\allowbreak\widetilde{h}_l)$, where
	\begin{itemize}
		\item $\widetilde{T}_l\subseteq T\times\mathbf{C_u}$ is the set of E$_l$SI-states associated with costs;
		\item $\Sigma^{pm}=((\Sigma_{\mathcal{I}}\cup\{\epsilon\})\times(\Sigma_1\cup\{\epsilon\})\times\dots\times(\Sigma_m\cup\{\epsilon\}))\setminus\{(\epsilon,\dots,\epsilon)\}$ is the set of $(m+1)$-tuple events and $\sigma^{pm}_{[i]}$, $i\in\{0,1,\dots,m\}$, denotes the $(i+1)$-st component of an event $\sigma^{pm}\in\Sigma^{pm}$;
		\item $\widetilde{T}_{l,0}=(\tau_l,0)\in \widetilde{T}_l$ is the initial state with zero cost;
		\item $\widetilde{T}_{l,e}\subseteq \{T_e\}\times\mathbf{C_u}$ is the set of marked and ending states associated with possible different costs;
		\item $\widetilde{h}_l:\widetilde{T}_l\times\Sigma^{pm}\rightarrow\widetilde{T}_l$ is the deterministic event release transition function, which is defined as follows: for any $(\tau,c)=(\tau^{(1)},\ldots,\tau^{(m)},c), (\tau',c')=(\tau'^{(1)},\ldots,\tau'^{(m)},c')\in \widetilde{T}_l$, $\sigma^{pm}\in\Sigma^{pm}$, it holds
		\begin{multline*}
			\widetilde{h}_l((\tau,c),\sigma^{pm})=(\tau',c')\Rightarrow\\((\sigma^{pm}_{[0]}=\epsilon\Rightarrow \text{LH1})\land(\sigma^{pm}_{[0]}\in\Sigma_{\mathcal{I}}\Rightarrow \text{LH2}))
		\end{multline*}
where 
\end{itemize}
		\begin{align*}
				\text{LH1} &\Leftrightarrow 
			\begin{aligned}[t]
				&(\exists i\in\mathcal{I},\exists(\epsilon,\sigma^{pm}_{[i]})\in E_{i,I}(\tau^{(i)}):\\
				\hspace{-1cm}& c'=c+[R_i]_{\epsilon,\sigma^{pm}_{[i]}}\leq c_u\land\tau'^{(i)}=\tau^{(i)}/\sigma^{pm}_{[i]})\land\\
				& (\forall j\in\mathcal{I}\setminus\{i\}:\sigma^{pm}_{[j]}=\epsilon\land\tau'^{(j)}=\tau^{(j)})
			\end{aligned}\\
			\text{LH2} &\Leftrightarrow\begin{aligned}[t]
&(\forall j\in\mathcal{I}\setminus I(\sigma^{pm}_{[0]}):\sigma^{pm}_{[j]}=\epsilon\land\tau'^{(j)}=\tau^{(j)})\land(\forall i\in\\ &I(\sigma^{pm}_{[0]}),\exists(\sigma^{pm}_{[0]},\sigma^{pm}_{[i]})\in \Phi_i(\tau^{(i)})\setminus E_{i,I}(\tau^{(i)}):
c'=\\&c+\Sigma_{i\in I(\sigma^{pm}_{[0]})}[R_i]_{\sigma^{pm}_{[0]},\sigma^{pm}_{[i]}}\leq c_u\land\tau'^{(i)}=\tau^{(i)}/\sigma^{pm}_{[i]}).
			\end{aligned}
		\end{align*}
\end{definition}

The domain of function $\widetilde{h}_l$ can be extended to $\widetilde{T}_l\times(\Sigma^{pm})^*$ in the standard recursive manner: $\widetilde{h}_l((\tau_{l1},c),\sigma^{pm}t^{pm})=\widetilde{h}_l(\widetilde{h}_l((\tau_{l1},c),\sigma^{pm}),t^{pm})$ for $(\tau_{l1},c)\in\widetilde{T}_l$, $\sigma^{pm}t^{pm}\in(\Sigma^{pm})^*$. 

In contrast to the E$_g$TS-builder, the event release transition function in an E$_l$TS-builder takes into account the release list $\Phi_i(\cdot)$ for each PO-sequence of every E$_l$SI-state and selects one feasible event to be the potential observation of the system, i.e., $\sigma^{pm}_{[0]}$ of a particular $\sigma^{pm}$. 
Furthermore, the error actions of insertions for each PO-sequence are always considered individually and separately from other error actions due to the following two reasons. 
First, similar to the discussion in Remark \ref{RE-OneInsertion}, we can always simulate the effects of several error actions of insertions in an E$_l$SI-state by considering consecutive single error actions of insertion. 
Second, unlike other error actions, this one does not offer potential observations in the system, indicating that it does not contribute to the update of the system state estimation. 
Therefore, given a transition $((\tau,c),\sigma^{pm},(\tau',c'))\in\widetilde{h}_l$, $\sigma^{pm}_{[0]}$ could be $\epsilon$ or not:

1) $\sigma^{pm}_{[0]}==\epsilon$: (LH1) holds if there exists $(\epsilon,\sigma^{pm}_{[i]})\in E_{i,I}(\tau^{(i)})$ that indicates the error action of insertion of the first symbol in $\tau^{(i)}$. 
Thus, only the first symbol in $\tau^{(i)}$ of $\tau$ is deleted, i.e., $\tau'^{(i)}=\tau^{(i)}/\sigma^{pm}_{[i]}$. 
Accordingly, if there exists $i\in\mathcal{\mathcal{I}}$ such that $|E_{i,I}(\tau^{(i)})|\neq0$, then for any $(\epsilon,\sigma)\in E_{i,I}(\tau^{(i)})$, $\widetilde{h}_l(\tau,(\epsilon,\epsilon,\dots,\sigma, \dots,\epsilon))!$ ($\sigma$ is in the $(i+1)$-st position).

2) $\sigma^{pm}_{[0]}\in\Sigma_{\mathcal{I}}$: (LH2) holds if, for all OSs whose sets of observable events contain $\sigma^{pm}_{[0]}$, each OS has a release list containing at least one pair of events with the possible original event being $\sigma^{pm}_{[0]}$, indicating that $\sigma^{pm}_{[0]}$ is a possible observation in the system. 
In other words, if there exists $i\in\mathcal{I}$, such that $\phi_i\in\Phi_i(\tau^{(i)})\setminus E_{i,I}(\tau^{(i)})$ and $\phi_i[0]=\sigma^{pm}_{[0]}$, then, for any $j\in I(\phi_i[0])$, there exists $\phi_j\in\Phi_j(\tau^{(j)})\setminus E_{j,I}(\tau^{(j)})$, such that $\phi_i[0]=\phi_j[0]$ ($\phi_i[0]$ and $\phi_j[0]$ are the first components of $\phi_i$ and $\phi_j$, respectively). 
Clearly, both cases have to satisfy the cost constraints. 

\begin{figure*}[htbp]
	\centering
	\includegraphics[scale=0.88]{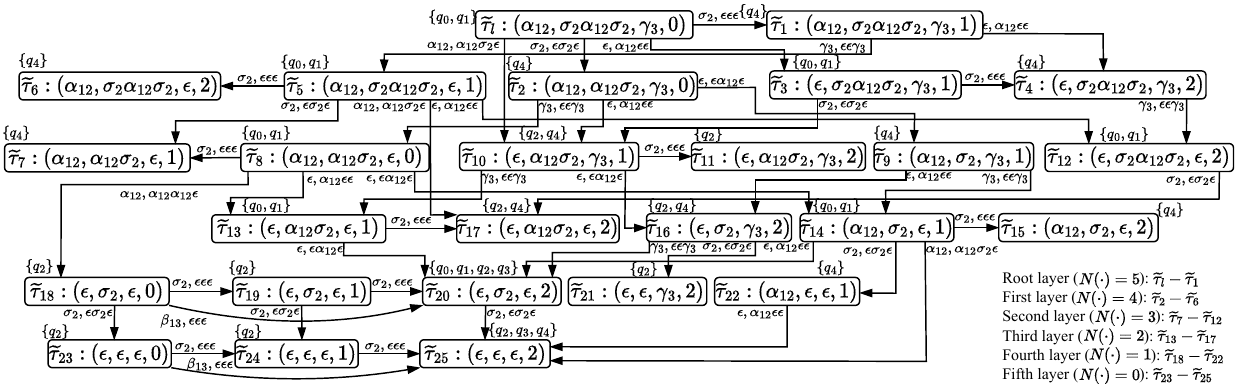}
	\caption{The E$_l$T-synchronizer of the system $G$ w.r.t. the given $\tau_{l}$ and $Q_0=\{q_0\}$ where $\operatorname{UR}(Q_0)=\{q_0,q_1\}$. The corresponding set of state estimates is displayed next to each E$_l$SI-state and the states in each layer are listed to the side.}
	\label{local-2-s}
\end{figure*}

\begin{example}
	Consider again the system in Example \ref{EX-local-1}. The sub-structure (this is explained later) of the E$_l$TS-builder given  $\widetilde{\tau}_l=(\alpha_{12},\sigma_2\alpha_{12}\sigma_2,\gamma_3,0)$ is illustrated in Fig. \ref{local-2-s}. We first examine the case when (LH1) holds in $\widetilde{\tau}_l$. Let $\tau_l=(\alpha_{12},\sigma_2\alpha_{12}\sigma_2,\gamma_3)$. 
	Since $[R_1]_{\epsilon,\alpha_{12}}=1\neq\infty$ and $(\epsilon,\alpha_{12})\in E_{i,I}(\tau_l^{(1)})$, $\alpha_{12}$ may be an inserted observation in $\tau_l^{(1)}$. Therefore, let $\widetilde{h}_l((\tau_l,0),(\epsilon, \alpha_{12},\epsilon,\epsilon))=\widetilde{\tau}_3$. Additionally, we discover that the first components of both $(\alpha_{12},\alpha_{12})\in\Phi_1(\tau_l^{(1)})\setminus E_{1,I}(\tau^{(1)})$ and $(\alpha_{12},\sigma_2)\in\Phi_2(\tau_l^{(2)})\setminus E_{2,I}(\tau^{(2)})$ share the identical symbol, $\alpha_{12}$. Hence, (LH2) holds due to $I(\alpha_{12})=\{1,2\}$ and $\widetilde{h}_l((\tau_l,0),(\alpha_{12}, \alpha_{12},\sigma_2,\epsilon))=\widetilde{\tau}_{10}$.\hfill\rule{1ex}{1ex}
\end{example}

We still use $t^{pm}_{[i]}$ to denote the sequence formed by the $i$-th element of each $m+1$-tuple in $t^{pm}$.
In this case, we have the following lemma.
The state space of an E$_l$TS-builder $\widetilde{B}_l$ is finite, for the same reasons previously explained for the finiteness of the E$_g$TS-builder.
\begin{lemma}\label{local-s-builder-string}
	Given an E$_l$SI-state $\tau_l$ w.r.t. $\{[R_i]\}^{c_u}_{i\in\mathcal{I}}$, and its corresponding E$_l$TS-builder $\widetilde{B}_l$, it holds:
	\begin{enumerate}
		\item $\forall t^{pm}\in L_m(\widetilde{B}_l):((T_e,c)=\widetilde{h}_l(\widetilde{T}_{l,0},t^{pm})\Rightarrow(\forall i\in\mathcal{I}:(t^{pm}_{[i]},c_i)\in\overline{\mathcal{R}}_i(P_i(t^{pm}_{[0]}))\land c=\Sigma^m_{i=1}c_i\leq c_u))$;
		\item $\mathcal{TO}^{c_u}_l(\tau_l)=\{(t^{pm}_{[0]},c)|\exists t^{pm}\in L_m(\widetilde{B}_l):(T_e,c)=\widetilde{h}_l(\widetilde{T}_{l,0},t^{pm})\}$.
	\end{enumerate}
\end{lemma}
The proof of the above lemma follows from Lemma \ref{global-GETO} and Definition \ref{E$_l$TS-builder}, and is therefore omitted here.
Lemma \ref{local-s-builder-string}.2) shows that the LETO-sequences for any $\tau_l$ can be obtained directly from its corresponding E$_l$TS-builder.
Similar to the analysis in global scenario, the DO-E$_{l}$SE is performed based on the E$_l$TS-builder, using the \textit{estimation-by-release} approach through the construction of the E$_l$T-synchronizer $\widetilde{\mathcal{S}}_l(\tau_l,Q_0)=(\widetilde{T}_l,\Sigma^{pm},\widetilde{T}_{l,0},\widetilde{T}_{l,e},\widetilde{h}_{l,s},\widetilde{c}_l,\widetilde{C}_l)$, which is detailed in Algorithm~\ref{E$_l$T-synchronizer}.
The construction process of the E$_l$T-synchronizer follows a similar approach to that used in Algorithms~\ref{E$_g$T-synchronizer} and~\ref{E$_l$-synchronizer}, utilizing a breadth-first search procedure. However, the release and estimation part in the construction of the E$_l$T-synchronizer combines the properties of both algorithms due to the following reasons: 
1) Given $(\tau,c)\in\widetilde{T}_l$, the release transitions originating at $(\tau,c)$ may end at $(\tau,c')$, i.e., at the same layer; 
2) Given $(\tau,c)\in\widetilde{T}_l$, the release transitions originating at $(\tau,c)$ may end at $(\tau',c')$ and $(\tau'',c'')$ within different subsequent layers. Taking these considerations into account, we design Algorithm \ref{E$_l$T-synchronizer} by integrating the methods proposed in the previous sections. During the breadth-first search, at each layer, we use DFS to compute the state estimation for E$_l$SI-states with the same $\tau$ but different cost components. Additionally, we use the function $N$ to label the E$_l$SI-state component in each $\widetilde{T}_l$ such that the state with the maximum number can be estimated first. 

\begin{algorithm}
	\caption{Construction of an E$_l$T-synchronizer}\label{E$_l$T-synchronizer}
	\begin{algorithmic}[1]
		\Require System $G$, set of states $Q_0$, E$_l$SI-state $\tau_l$, $\{[R_i]\}^{c_u}_{i\in\mathcal{I}}$.
		\Ensure $\widetilde{\mathcal{S}}_l(\tau_l,Q_0)=(\widetilde{T}_l,\Sigma^{pm},\widetilde{T}_{l,0},\widetilde{T}_{l,e},\widetilde{h}_{l,s},\widetilde{c}_l,\widetilde{C}_l).$
		\State $\widetilde{T}_{l,0}=(\tau_l,0)$, $\widetilde{T}_l=\{\widetilde{T}_{l,0}\}$, $\widetilde{c}_l(\widetilde{T}_{l,0})=\operatorname{UR}(Q_0)$, $\widetilde{C}_l=\{\widetilde{c}_l(\widetilde{T}_{l,0})\}$, Release Level $RL=\{N(\tau_l)\}$;
		\While{$RL\neq\emptyset$}
		\State \hspace{-0.3cm}
\textit{\textit{Estimation}}($\{(\tau,c)\in \widetilde{T}_l|N(\tau)=\max(RL)\}$);
		\State \hspace{-0.3cm}
$RL=RL\setminus\{\max(RL)\}$;
		\EndWhile
		\State \textbf{procedure} \textit{\textit{Estimation}}($\widetilde{T}_1$);
		\For{$\{(\tau,c)\in \widetilde{T}_1|\nexists(\tau,c')\in \widetilde{T}_1:c'<c\}$}
		\State \hspace{-0.3cm}
\textit{Esti\&DFS}($(\tau,c)$);
		\EndFor
		\State \textbf{procedure} \textit{\textit{Esti\&DFS}}($(\tau,c)$);
		\If{$(\tau,c)\neq(\tau_l,0)$} 
		\State 
\hspace{-0.3cm} $\widetilde{c}_l((\tau,c))=\bigcup_{((\tau',c'),\sigma^{pm},(\tau,c))\in\widetilde{h}_{l,s}}\operatorname{UR}(\operatorname{R}_{\sigma^{pm}_{[0]}}(\widetilde{c}_l(\tau',c')))$;
		\State 
\hspace{-0.3cm}
$\widetilde{C}_l=\widetilde{C}_l\cup\{\widetilde{c}_l((\tau,c))\}$;
		\EndIf
		\State \textit{Release}($(\tau,c)$);
		\If{$\exists(\tau,c_1):\widetilde{c}_l((\tau, c_1))=\emptyset$, such that $\nexists(\tau, c_2):c_2<c_1\land\widetilde{c}_l((\tau, c_2))=\emptyset$}
		\State \hspace{-0.3cm}
\textit{Esti\&DFS}($(\tau,c_1)$);
		\EndIf
		\State \textbf{procedure} \textit{Release}($(\tau,c)$)
		\For{$\sigma^{pm}\in\{\sigma^{pm}|\widetilde{h}_l((\tau,c),\sigma^{pm})!\land\operatorname{R}_{\sigma^{pm}_{[0]}}(\widetilde{c}_l((\tau,c)))\neq\emptyset\}$}
		\State \hspace{-0.3cm}
add $\widetilde{h}_l((\tau,c),\sigma^{pm})$ and $((\tau,c),\sigma^{pm},\widetilde{h}_l((\tau,c),\sigma^{pm}))$ to $\widetilde{T}_l$ and $\widetilde{h}_{l,s}$, respectively, $RL=RL\cup\{N(\widetilde{h}_l((\tau,c),\sigma^{pm}))\}$;
		\EndFor
	\end{algorithmic}
\end{algorithm}

\begin{theorem}\label{local-b-theorem}
	Given an E$_l$SI-state $\tau_l$ w.r.t. $\{[R_i]\}^{c_u}_{i\in\mathcal{I}}$, let $\widetilde{\mathcal{S}}_l(\tau_l,Q_0)=(\widetilde{T}_l,\Sigma^{pm},\widetilde{T}_{l,0},\widetilde{T}_{l,e},\widetilde{h}_{l,s},\widetilde{c}_l,\widetilde{C}_l)$ be the corresponding E$_l$T-synchronizer. It holds $\mathcal{E}^{c_u}_l(\tau_l,Q_0)=\{(q,c)|\exists(T_e,c)\in \widetilde{T}_{l,e}:q\in \widetilde{c}_l((T_e,c))\}$.
\end{theorem}

\begin{example}
	The E$_l$T-synchronizer w.r.t.  $\tau_l$ in Example \ref{EX-local-1} is shown in Fig. \ref{local-2-s}. Note that from $\widetilde{\tau}_l$, the DFS w.r.t the cost component is implemented to construct transition $(\widetilde{\tau}_l, (\sigma_2,\epsilon\epsilon\epsilon),\widetilde{\tau}_1)$ and assign state estimates to $\widetilde{\tau}_1$, such that $\widetilde{\tau}_l$ and $\widetilde{\tau}_1$ are in the same layer. At the same time, there exist also transitions $(\widetilde{\tau}_l, (\alpha_{12},\alpha_{12}\sigma_2\epsilon),\widetilde{\tau}_{10})$ and $(\widetilde{\tau}_l, (\sigma_2,\epsilon\sigma_2\epsilon),\widetilde{\tau}_2)$ released from $\widetilde{\tau}_l$ such that $\widetilde{\tau}_{10}$ and $\widetilde{\tau}_{2}$ are in the different layers. Thus, the E$_l$T-synchronizer possesses the properties of both E$_g$T-synchronizer and E$_l$-synchronizer. As seen, $\widetilde{T}_{l,e}=\{\widetilde{\tau}_{23},\widetilde{\tau}_{24},\widetilde{\tau}_{25}\}$ is the set of ending states and $\mathcal{E}^{c_u}_l(\tau_l,Q_0)=\{(q_2,0)\}\cup\{(q_2,1)\}\cup(\{q_2,q_3,q_4\}\times\{2\})=\{(q_2,0),(q_2,1),(q_2,2),(q_3,2),(q_4,2)\}$.~\hfill\rule{1ex}{1ex}
\end{example}

\section{Discussion and Comparison of two methods}

Given a system model $G=(Q,\Sigma,\delta,Q_0)$ and an ERM $[R]^{c_u}$, the modified system $G_g$ has at most $|Q||c_u+1|$ states. The transitions originating from a state with cost $(q,c)$ will not end at states $(q',c')$, such that $c>c'$. Hence, there are at most $\Sigma^{c_u+1}_{i=1}|Q||Q| i(|\Sigma_{\mathcal{I}}|+1)=\frac{1}{2}|Q|^2(c_u+1)(c_u+2)(|\Sigma_{\mathcal{I}}|+1)$ transitions (we assume that unobservable events, i.e., $\Sigma\setminus\Sigma_{\mathcal{I}}$, can be described as transition $\epsilon$). Thus, the worst-case time complexity for the construction of the modified system is 
$O(c_u^2|Q|^2|\Sigma_{\mathcal{I}}|)$. Similarly, the construction of $G_l$ can be done in $O(c_u^2|Q|^2\prod_{i=1}^{m}(|\Sigma_i|+1))$. 

In regard to the construction of the synchronizer, the complexity can be analyzed in two main components. 1) The first component involves the frequency of updating the state estimates, which corresponds to the transitions originating from a synchronizer state. 2) The second component pertains to set union operations, linked to the number of transitions that end at a synchronizer state. Consequently, the number of union operations within the construction of a synchronizer is bounded by the number of transitions. We use $\kappa_i$ to denote the length of PO-sequences reported by $O_i$. Then, the number of states of the corresponding S-builder is bounded by $\mathcal{K}=\prod_{i=1}^{m}(\kappa_i+1)$.

\textit{Complexity analyses of DO-E$_g$SE and DO-E$_l$SE of system modification (Method-A)}: The number of transitions of synchronizer $\mathcal{S}_g$ is bounded by $m\mathcal{K}$ (since an E$_g$SI-state can release at most $m$ symbols). The complexity of updating the state estimates associated to a transition is $O(\frac{1}{2}|Q|^2(c_u+1)(c_u+2))$ since, state $(q,c)$ can only be updated to states $(q',c')$ ($q'\in Q, c'\geq c$), which can be simplified to $O(c_u^2|Q|^2)$. The complexity of the union of sets of system states is $O(m\mathcal{K}|Q|(c_u+1))$. Therefore, the entire complexity of constructing synchronizer $\mathcal{S}_g$ is $O(m\mathcal{K}c_u^2|Q|^2+m\mathcal{K}|Q|(c_u+1))$ which can be simplified to $O(m\mathcal{K}c_u^2|Q|^2)$. In the scenario of local error, a state of the MS-builder has an upper bound of $(2^m-1)$ transitions originating from it. Therefore, the computational complexity of constructing the E$_l$-synchronizer is $O((2^m-1)\mathcal{K}\frac{1}{2}|Q|^2(c_u+1)(c_u+2)+(2^m-1)\mathcal{K}|Q|(c_u+1))$ which can be simplified to $O(2^m\mathcal{K}c_u^2|Q|^2)$.

\textit{Complexity analyses of DO-E$_g$SE and DO-E$_l$SE of S-builder modification (Method-B)}: 
The number of states in an E$_g$T-synchronizer is bounded by $\mathcal{K}(c_u+1)$. For an E$_g$SI-state with cost smaller than $c_u$, there are no more than $m(|\Sigma_{\mathcal{I}}|+1)+|\Sigma_{\mathcal{I}}|$ transitions originating from it, where $m(|\Sigma_{\mathcal{I}}|+1)$ is the number of transitions when H2 holds, and $|\Sigma_{\mathcal{I}}|$ is the number of transitions when H1 holds. For an E$_g$SI-state with cost $c_u$, the maximum number of transitions originating from it is $m$, corresponding to the error-less actions. Therefore,  an E$_g$T-synchronizer has at most $\mathcal{K}c_u(m(|\Sigma_{\mathcal{I}}|+1)+|\Sigma_{\mathcal{I}}|)+m\mathcal{K}$ transitions. However, each state only needs to compute state estimates at most $|\Sigma_{\mathcal{I}}|$ times (we also use this parameter to the case of E$_g$SI-state with cost $c_u$). Thus, the entire complexity of constructing an E$_g$T-synchronizer is
$O((c_u|\Sigma_{\mathcal{I}}|+|\Sigma_{\mathcal{I}}|)\mathcal{K}|Q|^2+(\mathcal{K}c_u(m(|\Sigma_{\mathcal{I}}|+1)+|\Sigma_{\mathcal{I}}|)+m\mathcal{K})|Q|
)$ which is simplified to $O(c_u\mathcal{K}|\Sigma_{\mathcal{I}}|(|Q|^2+m|Q|))$.

Similarly, in an E$_l$T-synchronizer, for any E$_l$SI-state with cost smaller than $c_u$, the number of transitions originating from it is bounded by $2^m(|\Sigma_{\mathcal{I}}|+1)$. For any E$_l$SI-state with the cost $c_u$, the number of transitions originating from it is at most $m$. Then, the number of transitions is bounded by $\mathcal{K}c_u2^m(|\Sigma_{\mathcal{I}}|+1)+m\mathcal{K}$. An E$_l$SI-state only needs to compute state estimates also at most $|\Sigma_{\mathcal{I}}|$ times. Hence, the worst-case time complexity of constructing an E$_l$T-synchronizer is $
O(
c_u|\Sigma_{\mathcal{I}}|+|\Sigma_{\mathcal{I}}|)\mathcal{K}|Q|^2
+
(\mathcal{K}c_u2^m(|\Sigma_{\mathcal{I}}|+1)+m\mathcal{K})|Q|
)
$, i.e., $O(c_u\mathcal{K}|\Sigma_{\mathcal{I}}|(|Q|^2+2^m|Q|))$.
\begin{table}[htbp]
	\centering
	\caption{Computation complexity comparison of approaches}\label{table.1}
\begin{tabular}{|m{0.7cm}|m{3.45cm}|m{3.4cm}|}
		\hline
		& \multicolumn{1}{c|}{Method-A}  &  \multicolumn{1}{c|}{Method-B}  \\
		\hline
		Global error & $O(c_u^2|Q|^2|\Sigma_{\mathcal{I}}|)+O(m\mathcal{K}c_u^2|Q|^2)$ & $O(c_u\mathcal{K}|\Sigma_{\mathcal{I}}|(|Q|^2+m|Q|))$ \\
		\hline
		Local error & $O(c_u^2|Q|^2\prod_{i=1}^{m}(|\Sigma_i|+1))+O(2^m\mathcal{K}c_u^2|Q|^2)$ & $O(c_u\mathcal{K}|\Sigma_{\mathcal{I}}|(|Q|^2+2^m|Q|))$ \\
		\hline
	\end{tabular}
\end{table}

In Table \ref{table.1}, the complexities of the methods explored in this study are concisely summarized. 
The first (second) components of the complexities for Method-A correspond to the complexities of modifying the system (performing state estimation).
Assuming that the modified system is precomputed, the state estimation procedures for both methods have polynomial complexities in $\mathcal{K}$ and $m$ under the global error, whereas under the local error, they have polynomial complexities in $2^m$ and $\mathcal{K}$ due to the nature of distributed setting.
In contrast to the error-free state estimation problem, whose complexity is of $O(m\mathcal{K}|Q|^2)$ \cite{SunHadjicostisLi2023},  the complexities of the proposed two methods depend on additional parameters concerning ERMs complexity, reflected by $c_u^2$ in Method-A, and by $c_u$ and $|\Sigma_{\mathcal{I}}|$ in Method-B, where $|\Sigma_{\mathcal{I}}|$ also reflects the scale of the system. 
Therefore, selecting an approach hinges on the intricacy of both the system at hand and the ERMs involved. 
It is noteworthy, however, that Method-A involves the establishment of system modification, thereby imposing a demand for supplementary computational resources. 
Imagine a scenario in which the ERM(s) or the bounded cost change between two consecutive synchronizations. 
In such instances, the modified systems must be repeatedly reconstructed, potentially leading to the consumption of computational resources. 
Conversely, Method-B is better equipped to navigate these challenges, offering a more resource-efficient solution.
Due to the nature of synchronizer, the computational complexities previously mentioned can be significantly mitigated depending on the number of shared-observable events and the time instants at which these events occur in the system evolution\cite{SunHadjicostisLi2023}. 
In addition, Method-B possesses the capability to infer potential system behavior, a feature that Method-A notably lacks.

\begin{remark}
The methods in this paper provide a foundation for analyzing security and privacy issues with applications including distributed control systems, supply chain management, and cyber-physical systems, especially those operating under the DO-based protocol.
The term ``error'' is used to describe inaccuracies in received information; however, it does not necessarily imply a deliberate attack.
In some cases, the ``error'' may result from an obfuscation method implemented by the system itself, similar to the insertion function described in \cite{JiWuLafortune2018}, which helps prevent intruders from determining sensitive information, such as the user's location or other private details.
The estimation approaches presented in this paper model the information accessible to an intruder to evaluate the effectiveness of this obfuscation technique.
\end{remark}

\section{Conclusion}

In this paper, we address current-state estimation in a discrete event system (DES), modeled as an nondeterministic finite automaton (NFA), under global/local errors. 
The system is observed at different observation sites and a coordinator utilizes the sequences of observations provided by the observation sites (OSs) to perform state estimation. 
Global errors (local errors) occur if a system observation is tampered before (after) being projected into partially ordered sequences of observations (PO-sequences). 
Error relation matrices (ERMs) are used to describe the possible errors upon each specific event. 
For each type of error, two methods are proposed to perform state estimation. 
The first method concerns system modification that can represent all possibly erroneous sequences of observations; the system is modified based on the given ERM and the synchronizer is constructed based on the breadth-first search (BFS) strategy. 
The second method involves inferring the matching original system  sequences of observations; a two-level hierarchical release strategy is used to construct the synchronizer due to the additional cost component in a synchronizer state compared to that in the first method. 
Thus, four different synchronizers are introduced to perform the tasks, each of which is constructed based on the \textit{estimation-by-release} approach.

\section*{Appendix}
\textit{Proof of Lemma \ref{G-A-Lemma}}
\begin{proof}
	($\subseteq$) 
	For any $(q,c)\in\mathcal{E}^{c_u}_g(\tau_g,Q_0)$, there exist $q_0\in Q_0$ and $u'\in L(G,q_0)$, such that $q\in\delta(q_0,u')$. 
	Moreover, letting $\omega=P_{\mathcal{I}}(u')$, there exists $(\omega_r,c)\in\overline{\mathcal{R}}^{c_u}(\omega)$ with $\omega_r\in\mathcal{TO}(\tau_g)$. 
	Suppose that $u'=u_1\sigma_1'u_2\sigma_2'\dots u_n\sigma_n'u_{n+1}$, where $u_i\in(\Sigma\setminus\Sigma_{\mathcal{I}})^*$, for $i\in\{1,\dots,n+1\}$ and $\sigma_j'\in\Sigma_{\mathcal{I}}$, for $j\in\{1,\dots,n\}$. 
	Then, $\omega=\sigma_1'\dots\sigma_n'$ and $\omega_r=s_1\sigma_1s_2\sigma_2\dots s_n\sigma_ns_{n+1}$, where $s_i\in\Sigma_{\mathcal{I}}^*$, for any $\alpha\in s_i$, $[R]_{\epsilon, \alpha}\neq\infty$, and $[R]_{\sigma'_j, \sigma_j}\neq\infty$, for $i\in\{1,\dots,n+1\}$ and $j\in\{1,\dots,n\}$. 
	Also, $c=\Sigma_{i\in\{1,\dots,n+1\}}\Sigma_{\alpha\in s_i}[R]_{\epsilon, \alpha}+\Sigma_{j\in\{1,\dots,n\}}[R]_{\sigma_j', \sigma_j}$. 
	Let $u=t_1\sigma_1t_2\dots t_i\sigma_it_{i+1}\dots t_n\sigma_nt_{n+1}$ where $t_i=u_{i_1}s_{i_1}\dots u_{i_j}s_{i_j}\dots u_{i_{n_i}}s_{i_{n_i}}u_{i_{n_i+1}}$ such that $u_i=u_{i_1}\dots u_{i_j}\dots u_{i_{n_i+1}}$ and $s_i=s_{i_1}\dots s_{i_k}\dots s_{i_{n_i}}$ for $i\in\{1,\dots,n+1\}$, $j\in\{1,\dots,n_i+1\}$, $k\in\{1,\dots,n_i\}$. 
	Moreover, $\omega_r=P_{\mathcal{I}}(u)$. 
	Since the events in $t_i$ either belong to $\Sigma\setminus\Sigma_{\mathcal{I}}$ or can be regarded as inserted events, according to the definition of $\delta_g$ in modified system $G_g$, $(q,c)\in\delta_g((q_0,0),u)$ holds, which completes the proof of $\mathcal{E}^{c_u}_g(\tau_g,Q_0)\subseteq\mathcal{E}^{G_g}(\tau_g,Q_0\times\{0\})$.	 
	
	The reverse $(\supseteq)$ can be proved similarly, which completes the proof of the lemma.	 	
\end{proof}

\textit{Proof of Theorem \ref{theorem-global-1}}
\begin{proof}
	First, we prove that for any $\tau\in T$, $c_g(\tau)=\{(q,c)\in Q\times\mathbf{C_u}|\exists t\in L(\mathcal{S},T_0):\tau=h_s(T_0,t)\land(\exists (q_0,0)\in Q_0\times \{0\},\exists u\in L(G_g,(q_0,0)):t=P_{\mathcal{I}}(u)\land (q,c)\in\delta((q_0,0),u))\}$.
	The proof is made by induction on the length of $t\in L(\mathcal{S},T_0)$.
	
	Induction basis: $|t|=0$. Based on the initial setup, for any $(q,c)\in c_g(T_0)=\operatorname{UR}(Q_0\times\{0\})$, there exist $(q_0,0)\in Q_0\times\{0\}$ and $u\in L(G_0,(q_0,0))$, such that $t=\epsilon=P_{\mathcal{I}}(u)$ and $(q,c)\in\delta((q_0,0),u)$. Then, the base case holds.
	
	Induction step: Assume that the induction hypothesis is true when $|t|=k$, $k\in\mathbb{N}$. Based on the definitions of the S-builder and $h_s$, it is evident that $h_s(\tau,e)$ is defined solely if event $e$ can be released in SI-state $\tau$ and align with permissible system behavior.

	We now consider the case $|t|=k+1$. Let $\tau=h_s(T_0,t)$. According to Algorithm 1 in \cite{SunHadjicostisLi2023}, we know that the assignment of state estimates to $\tau$ occurs only when the assignments of state estimates to $\tau$'s all preceding states are conducted, which is implemented via breadth-first search. Since $c_g(\tau_g)=\bigcup_{(\tau',e,\tau)\in h_s}\operatorname{UR}(\operatorname{R}(c_g(\tau')))$ (function $c_g$ collects all $\tau$'s preceding states and transitions and takes the union of their state estimates) and the induction hypothesis is true, combined with the definitions of operators $\operatorname{UR}$ and $\operatorname{R}$, the proof of ``$c_g(\tau)$'' part is explicitly completed. By substituting $T_e$ into $c_g(\tau)$ and in accordance with Lemma \ref{G-A-Lemma}, it follows that $\mathcal{E}^{c_u}_g(\tau_g, Q_0)=c_g(T_e)$, thereby completing the proof.
\end{proof}



\textit{Proof of Lemma \ref{global-GETO}}
\begin{proof}
	1) For any $t^p=\sigma_1^p\dots\sigma_i^p\dots\sigma_n^p\in L_m(\widetilde{B}_g)$, $i\in\{1,\dots,n\}$, if there exists $(T_e,c)\in\widetilde{T}_{g,e}$ such that $\widetilde{h}_g((\tau_g,0),t^p)=(T_e,c)$, according to the definition of $\widetilde{h}_g$, for any $\sigma_i^p\in t^p$, $[R]_{\sigma^p_{i[0]},\sigma^p_{i[1]}}\neq\infty$. 
	Thus, $(t^p[1],c)\in\overline{\mathcal{R}}(t^p[0])$, where $c=0+\Sigma_{i\in\{1,\dots,n\}}[R]_{\sigma^p_{i[0]},\sigma^p_{i[1]}}\leq c_u$. It leads to $(t^p[1],c)\in\overline{\mathcal{R}}^{c_u}(t^p[0])$.
	
	2) $(\subseteq)$ Given a state $\tau_g$, for any $(\omega,c)\in\mathcal{TO}^{c_u}_g(\tau_g)$, there exist $\omega_r\in\mathcal{TO}(\tau_g)$ and $(\omega_r,c)\in\overline{\mathcal{R}}^{c_u}(\omega)$. 
	Suppose that $\omega=S_1\sigma'_1S_2\dots S_i\sigma'_iS_{i+1}\dots\sigma'_nS_{n+1}$ and $\omega_r=\sigma_1\dots\sigma_i\dots\sigma_n$ where $S_j\in\Sigma_{\mathcal{I}}^*$; then, for any $\alpha_j\in S_j$, $[R]_{\alpha_j,\epsilon}\neq\infty$, $j\in\{1,\dots,n+1\}$, and $\sigma'_i\in \Sigma_{\mathcal{I}} \cup \{\epsilon\}$, $\sigma_i\in\Sigma_{\mathcal{I}}$, $[R]_{\sigma'_i,\sigma_i}\neq\infty$, $i\in\{1,\dots,n\}$ such that $c= \Sigma_{j \in \{1, \dots, n+1\}} \Sigma_{\alpha_j \in  S_j} [R]_{\alpha_j,\epsilon}+\Sigma_{i\in\{1,\dots,n\}}[R]_{\sigma'_i,\sigma_i}$.
	Let $t^p=t^{p}_1(\sigma'_1,\sigma_1)t^{p}_2\dots t^{p}_i(\sigma'_i,\sigma_i)t^{p}_{i+1}\dots t^{p}_n(\sigma'_n,\sigma_n)t^{p}_{n+1}$, where $t^p_j[0]=S_j$, and for any $\alpha^p_j\in t^p_j[1]$, $\alpha^p_j[1]=\epsilon$, $j\in\{1,\dots,n+1\}$. 
	Then, we have $\widetilde{h}_g(\widetilde{T}_{g,0},t^p)=(T_e,c)$ and $(t^p[1],c)\in\overline{\mathcal{R}}^{c_u}(t^p[0])]$ (based on Condition 1) of this lemma) such that $\omega=t^p[0]$ and $\omega_r=t^p[1]$,  which correctly indicates the ``$\subseteq$'' relation.
	
	$(\supseteq)$ This part of the proof follows from Condition~1) of this lemma which completes the proof of Condition~2).
\end{proof}

\textit{Proof of Theorem \ref{theorem-global-2}}

\begin{proof}
	First, we prove that for any $(\tau,c)\in\widetilde{T}_g$, we have
	\begin{multline*}
		\widetilde{c}_g((\tau,c))=\{q|\exists t^p\in L(\mathcal{\widetilde{S}}_g,\widetilde{T}_{g,0}): (\tau,c)=\widetilde{h}_{g,s}(\widetilde{T}_{g,0},t^p)\land\\ (\exists q_0\in Q_0,\exists u\in L(G,q_0):t^p[0]=P_{\mathcal{I}}(u)\land q\in\delta(q_0,u))\}.
	\end{multline*}
    The distinction between $\mathcal{S}_g$ and $\widetilde{\mathcal{S}}_g$ is manifested in the incorporation of an additional DFS procedure within each layer of $\widetilde{\mathcal{S}}_g$. Specifically, within each layer, we prioritize the E$_g$SI-state with the least cost for estimating the system states, thereby ensuring that both the release and the estimate rules are satisfied. Consequently, this proof follows from Theorem~\ref{theorem-global-1}.
	
	According to Lemma \ref{global-lemma-state} and Lemma \ref{global-GETO}.2), we have
	$\mathcal{E}^{c_u}_g(\tau_g,Q_0)=\{(q,c)|\exists t^p\in L_m(\widetilde{S}_g):(T_e,c)=\widetilde{h}_{g,s}(\widetilde{T}_{g,0}, t^p)\land(\exists q_0\in Q_0,\exists u\in L(G,q_0) :t^p[0]=P_{\mathcal{I}}(u)\land q\in\delta(q_0,u))\}$. Based on the aforementioned conclusion regarding $\widetilde{c}_g((\tau,c))$, we know that $\widetilde{c}_g((T_e,c))=\{q|\exists t^p\in L_m(\mathcal{\widetilde{S}}_g):(T_e,c)=\widetilde{h}_{g,s}(\widetilde{T}_{g,0}, t^p)\land(\exists q_0\in Q_0,\exists u\in L(G,q_0):t^p[0]=P_{\mathcal{I}}(u)\land q\in\delta(q_0,u))\}$. Combining these two equations, we conclude that $\mathcal{E}^{c_u}_g(\tau_g,Q_0) = \{(q,c)|\exists(T_e,c)\in\widetilde{T}_{g,e}:q\in \widetilde{c}_g((T_e,c))\}$.
\end{proof}

\textit{Proof of Lemma \ref{local-A-state}}
\begin{proof}
	$(\subseteq)$ For any $(q,c)\in\mathcal{E}^{c_u}_l(\tau_l,Q_0)$, there exist $q_0\in Q_0$ and $t=\sigma_1\dots\sigma_n\in L(G,q_0)$, such that $q\in\delta(q_0,t)$ and for any $i\in\mathcal{I}$, $(\tau^{(i)}_l,c_i)\in\overline{\mathcal{R}}_i(P_i(t))$ and $c=\Sigma^m_{i=1}c_i\leq c_u$. 
	We know that  $t^m=(\sigma_1^{(1)},\dots,\sigma_1^{(m)})\dots(\sigma_n^{(1)},\dots,\sigma_n^{(m)})$ where $\sigma^{(i)}_k= P_i(\sigma_k)$, $k\in\{1,\dots,n\}$.
	Then, for any $i\in\mathcal{I}$, $\tau^{(i)}_l=S^{(i)}_1e^{(i)}_1S^{(i)}_2\dots S^{(i)}_{n}e^{(i)}_{n}S^{(i)}_{n+1}$ with $S^{(i)}_j\in\Sigma_i^*$, such that for any $\alpha_j^{(i)}\in S^{(i)}_j$ and $j\in\{1,\dots,n+1\}$, $\alpha^{(i)}_j\in\overline{\mathcal{R}}_i(\epsilon)$; and $e^{(i)}_k\in\overline{\mathcal{R}}_i(\sigma^{(i)}_k)$ hold if $\sigma^{(i)}_k\neq\epsilon$ (otherwise, $e^{(i)}_k=\epsilon$), $k\in\{1,\dots,n\}$. 
	Also, $c_i=\Sigma_{j\in\{1,\dots,n+1\}}\Sigma_{\alpha_j^{(i)}\in S^{(i)}_j}[R_i]_{\epsilon,\alpha_j^{(i)}}+\Sigma_{k\in\{1,\dots,n\}\land\sigma^{(i)}_k\neq\epsilon}[R_i]_{\sigma^{(i)}_k,e^{(i)}_k}$ and $c=\Sigma^{m}_{i=1}c_i\leq c_u$. 
	Let $s^m=S_1(e^{(1)}_1,\dots,e^{(m)}_1)S_2\dots S_j(e^{(1)}_j,\dots,e^{(m)}_j)S_{j+1}\dots  S_n (e^{(1)}_n,\allowbreak\dots, e^{(m)}_n) S_{n+1}$ where $S_j\in(\Sigma^{m}_{l_{in}})^*$, $(\epsilon,\dots,\alpha^{(i)}_j,\dots,\epsilon)\in S_j$, $j\in\{1,\dots,n+1\}$, indicating the error action of insertion in the $i$-th component. 
	According to Remark \ref{RE-OneInsertion}, there is no specific ordering requirement between $(\epsilon,\dots,\alpha^{(i)}_j,\dots,\epsilon)$ and $(\epsilon,\dots,\alpha^{(k)}_j,\dots,\epsilon)$ in $S_j$. 
	By the definition of $\delta_l$ in the modified system $G_l$,  for any $i\in\mathcal{I}$, one has $l_i(s^m)=\tau^{(i)}_l$ and $(q,c)\in\delta_l((q_0,0),s^m)$, which correctly indicates the ``$\subseteq$'' relation.
	
	The reverse $(\supseteq)$ can be proved similarly, which completes the proof of the theorem.	
\end{proof}



%
%

\textit{Proof of Theorem \ref{local-b-theorem}}
\begin{proof}
	First, we need to prove that for any $(\tau,c)\in\widetilde{T}_l$,
	\begin{multline*}
		\widetilde{c}_l((\tau,c))=\{q|
		\exists t^{pm}\in L(\mathcal{\widetilde{S}}_l,
		\widetilde{T}_{l,0}):(\tau,c)=\widetilde{h}_{l,s}(\widetilde{T}_{l,0},t^{pm})\\\land
		(\exists q_0\in Q_0, \exists u\in L(G,q_0):t^{pm}[0]=P_{\mathcal{I}}(u)\land q\in\delta(q_0,u))
		\}.
	\end{multline*}
This proof follows from Theorem \ref{theorem-global-2}. Then, it is trivial that $\mathcal{E}^{c_u}_l(\tau_l,Q_0)=\{(q,c)|\exists(T_e,c)\in \widetilde{T}_{l,e}:q\in \widetilde{c}_l((T_e,c))\}$, which follows from Lemma \ref{local-lemma-state}, Lemma \ref{local-s-builder-string}.2) and Theorem~\ref{theorem-global-2}.
\end{proof}

\end{document}